\documentclass[11pt,a4paper]{article}
\usepackage{jheppub}
\usepackage[utf8]{inputenc}
\usepackage{enumitem}
\usepackage{bm}
\usepackage{bbm}
\usepackage{xcolor}
\usepackage{scalerel}
\usepackage{enumitem}  
\allowdisplaybreaks

\usepackage{wasysym}
\usepackage{tikz}
\usetikzlibrary{decorations.pathmorphing}
\tikzset{snake it/.style={decorate, decoration=snake}}
\usepackage{mathrsfs}
\newcommand*{\md}{{\cdot}}

\usepackage[english]{babel}

\usepackage{tikz-feynman}
\tikzfeynmanset{compat=1.1.0}
\tikzfeynmanset{
graviton/.style={circle, draw=green!60, fill=green!5, very thick, minimum size=7mm}
}
\tikzfeynmanset{
codot/.style={/tikz/shape=circle,
/tikz/fill=red,/tikz/minimum size=0.1cm,/tikz/inner sep=1.8pt}
}
\tikzfeynmanset{sb/.style=
{/tikz/shape=circle,
/tikz/fill=black,
 /tikz/minimum size=0.3cm,/tikz/inner sep=1.pt
 } }
\tikzfeynmanset{myblob/.style=
{/tikz/shape=rectangle,
/tikz/fill=red,
 /tikz/minimum size=0.2cm,
 } }
\tikzset{box/.pic={\filldraw[fill=black]  (0,0) circle (2.5pt);
				   \filldraw [fill=black] (0.5,0) circle (2.5pt);
			       \draw [line width=5pt] (0,0) -- (0.5,0);}}

\def\be{\begin{equation}}
\def\ee{\end{equation}}
\def\bea{\begin{eqnarray}}
\def\eea{\end{eqnarray}}
\def\beal{\begin{equation}\begin{aligned}}
\def\eeal{\end{aligned}\end{equation}}
\def\nn{\nonumber}

\def\eqn#1{eq.~\eqref{#1}}
\def\eqns#1#2{eqs.~\eqref{#1} and~\eqref{#2}}
\def\eqnss#1#2#3{eqs.~\eqref{#1}, \eqref{#2} and~\eqref{#3}}

\def\Tab#1{Table~{\ref{#1}}}

\def\sec#1{Section~{\ref{#1}}}

\def\app#1{Appendix~{\ref{#1}}}
\def\rcite#1{ref.~\cite{#1}}
\def\rcites#1{refs.~\cite{#1}}

\def\cL{\mathcal{L}}

\def\cO{\mathcal{O}}

\def\nn{\nonumber}
\def\be{\begin{equation}}
\def\ee{\end{equation}}
\def\bea{\begin{eqnarray}}
\def\eea{\end{eqnarray}}

\title{Kerr worldline--QFT action from Compton amplitude to infinite spin orders}

\author[a,b]{Maor Ben-Shahar,}
\author[c,b]{Lucile Cangemi,}
\author[b,d]{Henrik Johansson}

\affiliation[a]{Center for Theoretical Physics – a Leinweber Institute,
Massachusetts Institute of Technology, Cambridge, MA 02139, USA}
\affiliation[b]{Department of Physics and Astronomy, Uppsala University,
Box 516, 75120 Uppsala, Sweden}
\affiliation[c]{School of Mathematics and Maxwell Institute for Mathematical Sciences,
University of Edinburgh, EH9 3FD, UK}
\affiliation[d]{Nordita, Stockholm University and KTH Royal Institute of Technology, \\
Hannes Alfv\'{e}ns  v\"{a}g 12, 10691 Stockholm, Sweden}

\emailAdd{maorbs@mit.edu}
\emailAdd{lucile.cangemi@ed.ac.uk}
\emailAdd{henrik.johansson@physics.uu.se}

\preprint{
MIT-CTP/5991\\
\phantom{~} \hfill UUITP-38/25  \\
}

\abstract{
We develop a quadratic-in-Riemann worldline action for a Kerr black hole at infinite spin orders by matching to a proposed tree-level Kerr Compton amplitude, originally obtained from higher-spin QFT considerations. A worldline action is an effective theory, and as such the tree-level matching needs to be corrected by loop effects, including UV counter terms, renormalization, and higher-order matching to general relativity. However, we anticipate that many features of the Wilson coefficients of the proposed tree-level action will remain unchanged even after a loop-level matching. While the worldline action is given in closed form, it contains an infinite number of quadratic-in-Riemann operators $R^2$, even for the same-helicity sector. We argue that in the same-helicity sector the $R^2$ operators have no intrinsic meaning, as they merely remove unwanted terms produced by the linear-in-Riemann operators, which are well-established in the literature. The opposite-helicity sector is somewhat more complicated, it contains both $R^2$ operators that removes unwanted terms, and $R^2$ operators that add new needed terms to the Compton amplitude. We discuss and classify all independent $R^2$ operators that can feature in the worldline action.}

\begin{document}
\maketitle
\addtocontents{toc}{\protect\setcounter{tocdepth}{2}}

\section{Introduction}


The success of detecting gravitational waves from merging compact binaries~\cite{LIGOScientific:2016aoc,LIGOScientific:2017vwq}, along with the promising outlook for future experiments~\cite{Punturo:2010zz,LISA:2017pwj,Reitze:2019iox}, has over the last years driven a concerted effort toward more precise theoretical predictions of black hole dynamics~\cite{Buonanno:1998gg,Damour:2001tu,Blanchet:2013haa}. Recent emphasis on the post-Minkowskian (PM) expansion~\cite{Damour:2017zjx} has introduced new techniques inspired by quantum field theory (QFT) and scattering amplitudes~\cite{Bjerrum-Bohr:2018xdl,Cheung:2018wkq,Bern:2019nnu}. These methods have yielded novel scattering predictions for binary black holes from 3PM through 5PM orders~\cite{Bern:2019nnu,Bern:2019crd,Damour:2019lcq,Cheung:2020gyp,Kalin:2020fhe,Parra-Martinez:2020dzs,Bjerrum-Bohr:2021din,Brandhuber:2021eyq,Bern:2021dqo,Bern:2021yeh,Dlapa:2021npj,Dlapa:2021vgp,Jakobsen:2022psy,FebresCordero:2022jts,Khalil:2022ylj,Jakobsen:2023pvx,Jakobsen:2023hig,Damgaard:2023ttc,Dlapa:2024cje,Akpinar:2024meg, 
Driesse:2024xad,Bern:2024adl,Bini:2025vuk,Dlapa:2025biy, Bern:2025zno, Akpinar:2025bkt,Bern:2025wyd} and for higher-order waveforms~\cite{Herderschee:2023fxh,Brandhuber:2023hhy,Georgoudis:2023lgf,Georgoudis:2023ozp,Bohnenblust:2023qmy,Bini:2024rsy,Georgoudis:2024pdz,Bohnenblust:2025gir,Brunello:2025eso}.  A wide range of effective field theory (EFT) methods are now available for describing classical interactions and observables of gravitational physics. These include frameworks based on QFT scattering amplitudes, such as direct extraction of classical limits~\cite{Vaidya:2014kza,Cachazo:2017jef,Guevara:2017csg}, the KMOC approach~\cite{Maybee:2019jus,Cristofoli:2021vyo,Cristofoli:2021jas,Cristofoli:2022phh,Adamo:2022rmp}, the heavy-particle EFT framework~\cite{Damgaard:2019lfh,Aoude:2020onz,Aoude:2025xxq}, eikonal techniques~\cite{
DiVecchia:2023frv,KoemansCollado:2019ggb,DiVecchia:2021bdo,Heissenberg:2021tzo,Haddad:2021znf,Adamo:2021rfq,DiVecchia:2022piu,Bellazzini:2022wzv,Luna:2023uwd,Gatica:2023iws,Georgoudis:2023eke,Fernandes:2024xqr,Du:2024rkf}, double-copy constructions~\cite{Bern:2008qj,Bern:2010ue,Bern:2019prr,Adamo:2022dcm,Bern:2022wqg,Luna:2016due,Luna:2017dtq,Shen:2018ebu,Li:2018qap,Goldberger:2019xef,Plefka:2019wyg,Bautista:2019tdr,Kim:2019jwm,Monteiro:2020plf,Haddad:2020tvs,Carrasco:2020ywq,Carrasco:2021bmu,Brandhuber:2021kpo,Gonzo:2021drq,Shi:2021qsb,Almeida:2020mrg,Wang:2022ntx,CarrilloGonzalez:2022mxx,Johansson:2025grx}, and formulations relying on soft graviton theorems~\cite{Saha:2019tub,Manu:2020zxl,DiVecchia:2021ndb,Alessio:2022kwv,A:2022wsk,Alessio:2024wmz,Elkhidir:2024izo,Akhtar:2024lkk,Alessio:2024onn} or twistor space~\cite{Adamo:2020yzi,Adamo:2022mev,Adamo:2023fbj,Kim:2023vgb,Kim:2023aff,Kim:2024grz,Adamo:2025fqt}. 
Quantum effects such as Hawking radiation have also been explored recently~\cite{Goldberger:2020geb,Aoude:2024sve,Copinger:2024pai,Ilderton:2025umd,Aoki:2025ihc,Adamo:2025fqt,Ilderton:2025aql,Aoude:2025jvt,Carrasco:2025bgu}.
Other effective descriptions have been around for some time, such as standard worldline EFTs~\cite{Goldberger:2004jt,Goldberger:2007hy,Kol:2007bc,Goldberger:2009qd,Foffa:2013qca,Foffa:2016rgu,Kalin:2020mvi}, which can be promoted to include spinning degrees of freedom~\cite{Porto:2005ac,Porto:2006bt,
Porto:2008tb,Porto:2008jj,Levi:2008nh,Porto:2010tr,Porto:2010zg,
Levi:2010zu,Levi:2011eq,
Porto:2012as,Levi:2014gsa,
Levi:2014sba,Levi:2015msa,Levi:2015uxa,Levi:2015ixa, 
Levi:2016ofk,Maia:2017yok, 
Maia:2017gxn,
Levi:2020kvb,Levi:2020uwu,Liu:2021zxr,Bonocore:2025stf}. These and similar actions have been used in a worldline quantum field theory (WQFT) framework~\cite{
Mogull:2020sak,Jakobsen:2021smu, Jakobsen:2021lvp,Jakobsen:2021zvh,Comberiati:2022cpm,Ben-Shahar:2023djm,Damgaard:2023vnx,Haddad:2024ebn,Hoogeveen:2025tew,Mogull:2025cfn}, which is convenient for computing loop-level classical quantities. Computations are streamlined through worldline Feynman rules, or more recently using generalized unitarity~\cite{He:2025how,Haddad:2025cmw}, thus exploiting the quantum perspective.


Spin plays a central role in realistic descriptions of astrophysical rotating black holes and has also emerged as a fertile arena for theoretical exploration. By carefully analyzing scattering amplitudes of massive spinning particles, one can infer spin-induced multipole moments~\cite{Vaidya:2014kza}, encoded through the spin vector $S^\mu=m a^\mu$, or the Kerr ring-radius vector $a^\mu$. On general grounds, this should only work well if the quantum spin is approaching infinity  $s \to \infty$, since astrophysical black holes have enormous spin in Planck units.  However, Kerr black holes are very special classical systems that appear to often trivially interpolate between small and large spins; this property is called {\it spin universality}~\cite{Holstein:2008sx,Holstein:2008sw,Siemonsen:2019dsu}.\footnote{See \rcite{Cangemi:2022abk} for a rotating fundamental string that does not exhibit spin universality, instead its classical limit is inferred by non-trivial spin interpolation.} The effective cubic coupling of a Kerr black hole to a graviton was first understood to all orders in spin from the worldline action~\cite{Levi:2015msa} and subsequently from the exponential form of the stress energy tensor~\cite{Vines:2017hyw}. This was followed by the construction of an infinite family of spin-$s$ quantum amplitudes \cite{Arkani-Hamed:2017jhn}, subsequently shown to give the Kerr three-point amplitude in the classical limit to all spin orders~\cite{Guevara:2018wpp,Chung:2018kqs}. Using modern on-shell methods, these higher-spin amplitudes opened a new pathway for calculations at leading PM order~\cite{Guevara:2017csg,Guevara:2019fsj,Chung:2019duq} and beyond. Furthermore, it suggested a change of perspective in which Kerr black holes at weak coupling $G$ behave similar to point-like elementary objects~\cite{Siemonsen:2019dsu,Arkani-Hamed:2019ymq,Aoude:2020mlg,Cangemi:2022bew}.

Four-point Compton amplitudes, describing the interaction of a Kerr black hole with two gravitons, have been constructed from on-shell factorization for opposite-helicity~\cite{Arkani-Hamed:2017jhn} and same-helicity~\cite{Johansson:2019dnu} graviton configurations. However, the opposite-helicity Arkani-Hamed--Huang--Huang (AHH) amplitudes are known to possess spurious poles when extended to higher-spin states, $s> 2$, signaling unresolved ambiguities associated with contact terms. This issue has motivated a variety of approaches aimed at identifying the correct Compton contact interactions, drawing on criteria such as consistent high-energy behavior, conjectured symmetries,  structural principles or novel wordline actions~\cite{Chung:2018kqs,Falkowski:2020aso,Guevara:2020xjx,Aoude:2020mlg,Chen:2022clh,Bern:2022kto,Aoude:2022trd,Saketh:2022wap,Bjerrum-Bohr:2023jau,Bjerrum-Bohr:2023iey,Aoude:2023vdk,Haddad:2023ylx,Azevedo:2024rrf,Guevara:2024edh,Vazquez-Holm:2025ztz,Alessio:2025nzd,Bjerrum-Bohr:2025lpw,Akpinar:2025tct} or by direct comparison to general relativity (GR) using the Teukolsky equation~\cite{Bautista:2021wfy,Bautista:2022wjf,Bautista:2023sdf,Scheopner:2023rzp}. By contrast, for spins 
$s\le 2$, the spinning Compton amplitudes are well understood and can be obtained via the double-copy construction~\cite{Kawai:1985xq,Bern:2008qj,Bern:2010ue}, from corresponding gauge-theory amplitudes~\cite{Johansson:2019dnu,Bautista:2019evw,Johansson:2015oia,Chiodaroli:2021eug,Guevara:2021yud,Cangemi:2022bew} with $s\le 1$. These gauge-theory building blocks are commonly referred to as the root-Kerr theory~\cite{Arkani-Hamed:2019ymq}.

The QFT origins of the well-behaved AHH amplitudes were elucidated through an explicit analysis of their underlying elementary-particle Lagrangians in~\rcite{Chiodaroli:2021eug}. Using high-energy tree-level unitarity constraints also led to a unique $s=5/2$ Compton amplitude and QFT~\cite{Chiodaroli:2021eug}, at the price of breaking spin universality~\cite{Aoude:2022trd} and the double copy~\cite{Johansson:2019dnu}. Massive higher-spin gauge symmetry~\cite{Zinoviev:2001dt} allows one to consistently impose good high-energy behavior and correct degrees of freedom (needed for unitarity), and in~\rcite{Cangemi:2022bew} it was shown that all previously known Kerr black hole amplitudes were uniquely fixed by imposing this symmetry. However, for $s\ge 3$ further constraints are needed. In~\rcite{Cangemi:2023ysz}, the higher-spin symmetry was supplemented by a chiral-field formulation~\cite{Ochirov:2022nqz} and the observation that symmetric homogeneous polynomials (generalization of geometric sum) play a crucial role in Kerr and root-Kerr amplitudes. This motivated a conjectured closed-form expression for the entire family of spin-$s$ Kerr Compton amplitudes~\cite{Cangemi:2023bpe}.

After taking an appropriate classical limit of the quantum Compton amplitude -- either via the infinite-spin limit~\cite{Cangemi:2022abk,Cangemi:2023ysz} or through the use of coherent spin states~\cite{Aoude:2021oqj}, which both works for $s>2$ when spin universality is absent -- leads to a well-behaved classical Compton amplitude to all orders in spin~\cite{Cangemi:2023bpe}. Matching this amplitude to direct GR computations using black-hole perturbation theory and the Teukolsky formalism~\cite{Bautista:2022wjf,privateBGKV,Bautista:2023sdf}, gives agreement without finetuning, up to certain non-rational polygamma functions (specifically, for the choice $\alpha=0$ of the bookkeeping parameter from~\rcite{Bautista:2022wjf}). In general, one would associate rational functions with tree level, and non-rational with loop level; however, such a splitting is not mathematically unique. This necessitates further studies where loop-level EFT matching to the Teukolsky results would clarify the situation. Dissipative effects -- labeled by $\eta$ -- also appear in the Teukolsky results~\cite{Bautista:2022wjf,privateBGKV}; these can be consistently incorporated into the higher-spin QFT Compton amplitude~\cite{Cangemi:2023bpe}; however, it is again unclear whether they should be considered tree-level effects~\cite{Bautista:2022wjf} or loop effects~\cite{Bautista:2023sdf} in an EFT framework. Specifically, in the week-coupling limit (super-extremal Kerr) these terms become odd in the spin magnitude $|a|$ and also odd in the graviton energy $\omega$, which makes them more complicated to deal with compared to conservative interactions. 

There is by now an extensive body of work using, or motivated by, amplitudes techniques applied to spinning compact objects, which have been studied from 2PM through 4PM~\cite{Guevara:2018wpp,Bini:2018ywr,Guevara:2019fsj,Chung:2019duq,Bern:2020buy,Chung:2020rrz,Chen:2021kxt,Kosmopoulos:2021zoq,Jakobsen:2022fcj,Damgaard:2022jem,Aoude:2022thd,FebresCordero:2022jts,Menezes:2022tcs,Jakobsen:2022zsx,Jakobsen:2022zsx,Bautista:2023szu,Chen:2024mmm,Akpinar:2024meg,Jakobsen:2023ndj, Bohnenblust:2024hkw,Buonanno:2024vkx,Gonzo:2024zxo,Akpinar:2025bkt} at various spin-multipole orders. Likewise, waveforms from the scattering of spinning binary black holes, as computed from five-point amplitudes of spinning particles, have been explored up to $\cO(G^3)$ order~\cite{Jakobsen:2021lvp,Bautista:2021inx,Alessio:2022kwv,Riva:2022fru,Heissenberg:2023uvo,DeAngelis:2023lvf,Brandhuber:2023hhl,Aoude:2023dui,Bohnenblust:2023qmy}. While still in an early phase, spin-magnitude and mass transitions and their relation to absorptive scattering have been studied from a QFT framework in \rcites{Aoude:2023fdm,Jones:2023ugm,Chen:2023qzo,Bautista:2024emt,Gatica:2025uhx}. See also recent work on conservative spin-magnitude change~\cite{Bern:2023ity,Kim:2023drc,Alaverdian:2024spu,Alaverdian:2025jtw}. Besides the above discussed work on Compton amplitudes, it is worth highlighting~\rcites{Bjerrum-Bohr:2023jau,Bjerrum-Bohr:2023iey} that give a closed-form tree-level Compton amplitude with the same classical factorization poles as the higher-spin proposal~\rcite{Cangemi:2023bpe}, while the contact terms are different. Compton amplitudes for scalars and photons in Kerr background have also been put forward in~\rcite{Correia:2024jgr} to high orders in spin, and recently several calculations of one-loop Compton amplitudes have been obtained in~\rcites{Caron-Huot:2025tlq,Bjerrum-Bohr:2025bqg,Akpinar:2025byi,Akpinar:2025huz}.

In this paper, we construct a novel spining worldline action aimed at describing a Kerr black hole, with worldline operators up to quadratic order in the Riemann tensor and infinite order in the spin multipole expansion. The Riemann-square operators serves two purposes: removing unwanted terms that the linear-in-Riemann worldline~\cite{Levi:2015msa} produces in the Compton amplitude, and adding new terms in order to reproduce the tree-level opposite-helicity Compton amplitude proposed in~\rcite{Cangemi:2023bpe}. We discuss how the linear-in-Riemann worldline does not reproduce the expected exponential amplitude $\sim e^{a\cdot q}$ in the same-helicity sector~\cite{Johansson:2019dnu,Aoude:2020onz}, which gives rise to an infinite set of countering Riemann-squared operators and Wilson coefficients in the action that do not encode any physical information (since they leave no imprint on the Compton amplitude). Similar artifacts appear in the opposite-helicity amplitude, and we identify the infinite number of terms in the action, called $\frak{B_i}$, that do not carry any physical information. 

We also streamline and classify all possible Riemann-square operators that can be added to the worldline action, including dissipative operators which we treat using the Schwinger-Keldysh approach. We find that the counting of operators agree with \rcite{Haddad:2023ylx}. As part of the classification, we distinguish between operators that {\it vanish} versus {\it contribute} to polar scattering kinematics, which is an important characteristic for identifying the interactions of a Kerr black hole, as originally emphasized in \rcite{Bautista:2022wjf} and further elaborated on in \rcite{Cangemi:2023bpe}. 

The paper is structured as follows: In section \ref{sec_two}, we begin reviewing the construction of a spinning worldline effective action and how to use it for the computation of Compton scattering amplitudes. In section \ref{Sect3}, we present the all-orders-in-spin Compton amplitude obtained from the naive worldline with only linear-in-Riemann operators in the action. Using the same variables and pole structures as in \rcite{Cangemi:2023bpe} drastically simplifies the final result and allows us to write both the same- and opposite-helicity sector amplitudes in a compact form using entire functions. In section \ref{sect:MatchingR^2}, we compare the worldline results to the ones obtained from the higher-spin QFT amplitudes of \rcite{Cangemi:2023bpe}, and uplift the contact-term differences between them to $R^2$ operators in the worldline action. Both the contact terms difference and the operator are compactly written on closed form as entire functions, the latter using differential operators acting on the curvature tensors of the action. The conclusions are in section~\ref{sec_conclusion}. In appendices \ref{appendix_A} and \ref{appendix_action_to_onshell}, we spell out some conventions and how $C$, $P$, $T$ acts on the fields and variables, and elaborate on the mapping between worldine operators and on-shell variables.

\section{Worldline action for a Kerr black hole}
\label{sec_two}
We begin by reviewing a standard worldline EFT action for a dynamical Kerr black hole. Spin magnitude and mass are assumed to be conserved quantities, unless otherwise stated. 

\subsection{Minimal and non-minimal terms}
The EFT for a compact spinning massive body coupled to gravity can be described by the worldline action \cite{Levi:2015msa}
\begin{align} \label{ppEFT}
    S = &-\int d\tau \bigg(
    p_\mu \dot{x}^\mu + \frac{1}{2}S_{\mu\nu}\Omega^{\mu\nu} + 
    \frac{D {p}_\mu}{d\tau} S^{\mu\nu}\frac{p_\nu}{p^2}
    -\frac{\ell}{2}(p^2-\mathcal{M}^2) - \ell^\mu S_{\mu\nu}\Big(\frac{p^\nu}{|p|} + \Lambda_0^\nu\Big)
    \bigg)\,,
\end{align} 
where the fundamental worldline fields (independent dynamical variables) are the coordinates $x^\mu$, linear momentum $p^\mu$, spin tensor $S^{\mu \nu}$, and frame fields $\Lambda_I^\mu$, where the $I$ are flat Lorentzian indices. The angular velocity tensor is defined in terms of the frame fields $\Omega^{\mu\nu} = \Lambda^\mu_I\frac{D}{d\tau}\Lambda^{\nu I}$. The covariant derivative along the worldline acts as $\frac{D}{d\tau}p^\mu=\dot{p}^\mu+ \dot{x}^\rho\Gamma_{\rho\nu}^\mu p^\nu$, where the dot is the derivative $\frac{d}{d\tau}$. We use $|p|=\sqrt{p^2}$, and two Lagrange multipliers $\ell, \ell^{\mu}$ that enforce the mass-shell constraint and a spin-tensor constraint, the latter ensuring that only three degrees of freedom propagate in the spin sector. The dynamical mass $\mathcal{M}$ has an expansion in powers of Riemann tensors,
\begin{equation} \label{eq:dynamicalMass}
    \mathcal{M}^2 = m^2 + \mathcal{L}_{R}+ \mathcal{L}_{R^2} + \mathcal{O}(R^3)\,,
\end{equation}
where in this paper we will focus on the linear and quadratic curvature terms. 
The acceleration term can be removed by use of the equations of motion, 
\begin{equation}
    \frac{D p_\mu}{d\tau} = \frac{1}{2}\dot{x}^\rho R_{\mu\rho\nu\sigma}S^{\nu\sigma}+\frac{\ell}{2}D_\mu \mathcal{M}^2 
\end{equation}
but this introduces additional couplings to the curvature so we do not make use of this replacement here.
In addition to the worldline action \eqref{ppEFT}, the pure gravity interactions are described by the Einstein-Hilbert action,
\begin{equation} \label{eq:eh}
    S_{\mathrm{EH}} = -\frac{2}{\kappa^2}\int d^4 x \sqrt{-g} R\, .
\end{equation}

Let us now briefly summarize the gauge-fixing procedure for the worldline fields following ref.~\cite{Ben-Shahar:2023djm}. The minimal worldline action, defined by $\mathcal{M}^2 = m^2$, enjoys the spin-gauge symmetry
\begin{align}
    \delta S_{\mu\nu} &=\frac{2}{|p|}p_{[\mu}S_{\nu]\alpha}\epsilon^\alpha\, , \nn \\
    \delta\Lambda_I^\mu &= \frac{2}{|p|}\epsilon^{[\mu}p^{\nu]}\Lambda_{I\nu} + 2\epsilon^{[\mu}\Lambda_0^{\nu]}\Lambda_{I\nu} \,, \\
    \delta \ell^\mu &= -\frac{D\epsilon^\mu}{d\tau} + \ldots\,, \nn  
\end{align}
where the ellipsis corresponds to additional shifts in the Lagrange multiplier $\ell^\mu$ needed to keep the action invariant. Since $\ell^\mu$ is arbitrary, these can be freely chosen to cancel terms in the variation of the action, which are proportional to the constraint itself.
The gauge symmetry can be used to fix the vector Lagrange multiplier to
\begin{align}
\ell_\mu = \frac{1}{|p|}\frac{Dp_\mu}{d\tau} \, ,
\end{align}
such that a simplified worldline action is obtained \cite{Ben-Shahar:2023djm},
\begin{align} \label{ppEFT_gfixed}
    S = -\int d\tau \Big(
    p_\mu \dot{x}^\mu + \frac{1}{2}S_{\mu\nu}\Omega^{\mu\nu} 
    -\frac{\ell}{2}(p^2-\mathcal{M}^2) - \frac{1}{|p|}\frac{D p^\mu}{d\tau} S_{\mu\nu}\Lambda_0^\nu
    \Big) \, .
\end{align}
Additionally, the worldline has the usual gauge symmetry corresponding to time reparametrization invariance, which can be fixed by setting the scalar Lagrange multiplier to $\ell=m^{-1} $.

In order to ensure that the remaining terms in the action are gauge invariant, in the dynamical mass ${\cal M}$ we make use of the spin vector 
\begin{equation} \label{defSvector}
    S^\mu  = \epsilon^{\mu\nu\rho\sigma}\hat{p}_\nu S_{\rho\sigma} \, ,
\end{equation}
where we introduced the normalized momentum vector 
\begin{align}\label{P1_prescription}
\hat{p}^\mu &= \frac{p^\mu}{\sqrt{p^2}} \, .
\end{align}
The curvature operators~\eqref{eq:dynamicalMass} are then functionals of the gauge-invariant worldline fields and covariant derivatives,\footnote{Note that there is considerable freedom in choosing which worldline fields enter the $\mathcal{L}_{R^n}$.
For example, one can trade $p$ for velocity $\dot x$, and trade $p\cdot \nabla$ for the covariantized time derivative $\frac{D}{d\tau}$.}
\be
\mathcal{L}_{R^n} =\mathcal{L}_{R^n}(p_\mu,S^\mu,\nabla_\mu, E_{\mu\nu},B_{\mu\nu})\,.
\ee
Here we have decomposed the Riemann tensor into fields involving the electric and magnetic components that are transverse to the worldline momentum,
\begin{align} \label{EBfields}
E_{\mu \nu} &:= \hat{p}^\rho \hat{p}^\sigma R_{\mu \rho \nu \sigma} \, , \nn \\
B_{\mu \nu} &:= \frac{1}{2}\hat{p}^\rho \hat{p}^\sigma {\epsilon_{\nu\sigma}}^{ \kappa \lambda}  R_{\mu \rho \kappa \lambda} \, .
\end{align}
It is well known~\cite{Levi:2015msa} that a complete basis (up to terms that vanish by equations of motion) of parity-invariant linear-in-Riemann operators   for a compact spinning massive body can be written as
\begin{align}
\mathcal{L}_{R} =& 
       2\sum_{j=1}^\infty \frac{(-1)^{j+1}}{(2j)! m^{2j-2}}c_{ES^{2j}}\nabla_{\mu_1}\cdots \nabla_{\mu_{2j-2}}{E}_{\mu_{2j-1}\mu_{2j}}S^{\mu_1}\cdots S^{\mu_{2j}} \nn
        \\
        &+2\sum_{j=1}^\infty \frac{(-1)^{j}}{(2j+1)! m^{2j-1}}c_{BS^{2j+1}}\nabla_{\mu_1}\cdots \nabla_{\mu_{2j-1}}{B}_{\mu_{2j}\mu_{2j+1}}S^{\mu_1}\cdots S^{\mu_{2j+1}}  \, ,
\end{align}
where $c_{ES^{2j}}$, $c_{BS^{2j+1}}$ are the Wilson coefficients of the spin-multipole expansion.

For a Kerr black hole, the Wilson coefficients are all set to unity~\cite{Levi:2015msa}, $c_{ES^{2j}}=c_{BS^{2j+1}}=1$, and the operators can then be re-summed into an entire functional,
\begin{equation} \label{NonMinRop}
\mathcal{L}_{R} = \frac{2m^2}{(S\md \nabla)^2}\Big[\Big(1-e^{\frac{i}{m} S\md\nabla}+\frac{i}{m}  S\md\nabla\Big)E^+_{SS}
    + \Big(1-e^{-\frac{i}{m} S\md\nabla}-\frac{i}{m} S\md\nabla\Big)E^-_{SS}\Big]\,.
\end{equation}
Here we switched to a more convenient basis for the electric and magnetic tensors, by using selfdual and anti-selfdual curvature fields, 
\begin{align}
E^{\pm}_{\mu \nu}:= \frac{1}{2}(E_{\mu \nu} \pm i B_{\mu \nu})\,,
\end{align}
and also employed Schoonship notation $E^{\pm}_{S S}:=E^{\pm}_{\mu \nu} S^\mu S^\nu$. An equivalent form of \eqn{NonMinRop} but with electric and magnetic curvature tensors, is given in eq.~(7.48) of \rcite{Scheopner:2023rzp}.

In addition to the given non-minimal curvature operators \eqref{NonMinRop}, we will in parallel consider a slightly different prescription that removes worldline fields in the denominators. Specifically, our linear-in-Riemann curvature operators contain a proliferation of $1/|p|$ factors that come from the normalized momentum $\hat{p}$ in \eqns{defSvector}{EBfields}. We can remove this by using the alternative prescription\footnote{The minimal action \eqref{ppEFT}, or \eqref{ppEFT_gfixed}, also has a $1/|p|$ term; however, we do not modify it since this would spoil the gauge symmetry. } where the momentum is normalized by the mass,
\begin{align}\label{P2_prescription}
{\mathcal{L}_{R}}':={\mathcal{L}_{R}}\Big|_{\hat{p}^\mu \to  p^\mu/m} \, . 
\end{align}
On shell this is equivalent to the previous definition, and thus this prescription will give different interactions starting at quadratic order in the curvature. This corresponds to a contact-term difference, which can be worked out explicitly using the equation of motion $p^2=\mathcal{M}^2$,
\begin{equation} \label{InversePpowers}
    \frac{1}{|p|^n}= \frac{1}{\sqrt{m^2 + \mathcal{L}_R + \mathcal{O}(R^2)}^n}  = \frac{1}{m^n} -   \frac{n}{2m^{n+2}} \mathcal{L}_R + \mathcal{O}(R^2) \, .
\end{equation}
Thus, the difference between the two prescriptions is given by the quadratic curvature term
\begin{equation}  
\mathcal{L}_{R}-{\mathcal{L}_{R}}'= -\frac{\mathcal{L}_{R} 
{\widetilde{\mathcal{L}}}_{R}}{m^2}+\mathcal{O}(R^3)\,,
\end{equation}
where the ${\widetilde{\mathcal{L}}}_{R}$ corresponds to inserting appropriate integer factors $n$, as per \eqn{InversePpowers}, for each inverse power of momenta, which is correlated with the spin multipole, giving
\begin{equation}    
{\widetilde{\mathcal{L}}}_{R} := \frac{1}{2}\mathcal{L}_{R}\Big|_{S^n\to (n+2)S^n} = \frac{m^2}{(S\md \nabla)^2}\sum_{\pm}\Big[2-\Big(2\pm\frac{i}{m}  S\md\nabla\Big)e^{\pm\frac{i}{m} S\md\nabla}\pm\frac{3i}{m}  S\md\nabla\Big]E^\pm_{SS}\,.
\end{equation}

Thus, while the cubic actions for the two prescriptions produce the same three-point amplitudes, the corresponding Compton amplitudes differ due to the presence of contact terms coming from the quadratic in Riemann term $\mathcal{L}_{R} 
{\widetilde{\mathcal{L}}}_{R}$. We will use both prescriptions in this paper, since both have pros and cons. As for the remaining quadratic curvature corrections ${\cal L}_{R^2}$, we will work them out in section~\ref{sect:MatchingR^2}, by matching to the proposed all-orders-in-spin Compton amplitude of ref.~\cite{Cangemi:2023bpe}.

To be clear about the precise worldline action that we use in \sec{Sect3}, which contains only up to linear-in-Riemann terms, we quote it here: 
\begin{align} \label{ppEFT_gfixed_linearR}
    S = -\int d\tau \Big(
    p_\mu \dot{x}^\mu + \frac{1}{2}S_{\mu\nu}\Omega^{\mu\nu} 
    -\frac{\ell}{2}(p^2-m^2-\beta{\cal L}_R-(1-\beta){\cal L}_R') - \frac{1}{|p|}\frac{D p^\mu}{d\tau} S_{\mu\nu}\Lambda_0^\nu
    \Big) \, ,
\end{align}
the $\beta$ is an auxiliary parameter that lets us interpolate between the two prescriptions $\beta=0$ and $\beta=1$.

\subsection{Perturbation theory}

In the WQFT approach, Feynman rules are used to compute contributions to the amplitude. For the spinning particle, we decompose the frame fields $\Lambda_I^\mu$ by introducing a tetrad $e^a_\mu$ such that $\Lambda^\mu_I=\Lambda^a_Ie_a^\mu$. 
In the scattering scenario, we identify the fundamental spin fields of the worldline, $S_{ab}$ and $\Lambda_{a}^{I}$, with respect to this flat frame. The fields $\Lambda_{a}^{I}(\tau)$ are the  Lorentz matrices defined on the worldline and parametrize how the body-fixed frame at time $\tau$ deviates from the body-fixed frame in the asymptotic past. The tetrad $e_{\mu}^a$ is used to define the fields in curved space such that $S_{\mu\nu} = e^{a}_{\mu}e^{b}_{\nu}S_{ab}$. From now on, we will expand around flat space:  $e_\mu^a=\delta_\mu^a+\frac{\kappa}{2}h_\mu^a +\ldots$, subject to the constraint $g_{\mu \nu}=e_\mu^a e_{\nu a}=\eta_{\mu\nu}+\kappa h_{\mu\nu}$, such that the distinction between flat and curved indices is lost. As a consequence, we choose to work with Greek letters to denote flat indices in the perturbative calculations in the remaining parts of this paper.

Perturbing around the linearized solution is then achieved with the following substitutions into the action: 
\begin{align}
p_\mu &\to mv_\mu + \pi_\mu \, , \nn \\
x^\mu &\to b^\mu + v^\mu \tau + z^\mu \, , \nn\\
S_{\mu\nu} &\to S_{\mu\nu} + s_{\mu\nu} \, ,  \\
\Lambda^\mu_I &\to\Lambda_I^\mu + \lambda^{\mu\nu}\Lambda_{I\nu} + \frac{1}{2}\lambda^{\mu\nu}\lambda_{\nu\rho}\Lambda^\rho_I +\ldots \, ,  \nn
\end{align}
where the perturbations constitute the new set of dynamical fields 
$\{\pi_\mu,z^\mu,s_{\mu\nu},\lambda_{\mu\nu}\}$, and the remaining terms are background fields obeying the flat-space equations of motion $\dot{S} = \dot{\Lambda}=0$ and $v^2=1$. 
In the rest of the paper, we use ${S}_{\mu\nu}$ and $\Lambda_I^\mu$ to only denote the constant solutions to the equations of motion. The field $\lambda_{\mu\nu}$ is an antisymmetric matrix introduced through the exponential map of an infinitesimal rotation acting on the $\Lambda_I^\mu$ fields. In addition, it is convenient to take the asymptotic states to obey the covariant spin supplementary condition (covariant SSC), meaning we apply the gauge choices $\Lambda^\mu_0 = v^\mu$ and $S_{\mu\nu}v^\nu = 0$ on the constant spin tensors. 

In order to compute the worldline propagators, we can now expand the action up to quadratic order in the worldline perturbations, giving the kinetic terms
\begin{align}
S_{\mathrm{kin}}=& -\!\int d\tau\Big(
\dot{z}^\mu\pi_\mu
- \frac{1}{2m}\pi^2
+ \frac{1}{2}S^{\mu\nu}\lambda_{\mu\rho} \dot{\lambda}^{\nu\rho}
- \frac{1}{2}\dot{\lambda}^{\mu\nu}s_{\mu\nu} 
- \frac{1}{m} S^{\mu\nu}\dot{\pi}_\mu\lambda_{\nu\rho}v^\rho
- \frac{1}{m}\dot{\pi}^\mu s_{\mu\nu}v^\nu
\Big)\,.
\end{align}
Working in momentum/frequency space implies the insertion of Fourier transforms, e.g.
\begin{align}
z^\mu(\tau) = \int \frac{d\omega}{2\pi} e^{i\omega t}z^\mu(\omega) \, , 
\end{align}
with similar transforms for the other worldline perturbations (we work with all-incoming momenta and frequencies $\omega$).
Inverting the kinetic terms, gives the following non-vanishing two-point functions or propagators:
\begin{align}\label{eqn:propagators}
	\big\langle z^\mu(-\omega) z^\nu(\omega)\big\rangle =& -i\frac{1}{m\omega^2}\eta^{\mu\nu} - \frac{1}{m^2 \omega} {S}^{\mu\nu}\, , \nonumber \\
	\big\langle \pi^\mu(-\omega) z^\nu(\omega)\big\rangle =& -\frac{1}{\omega}\eta^{\mu\nu}\, , \nonumber \\
	\big\langle s_{\mu\nu}(-\omega)s_{\rho\sigma}(\omega) \big\rangle =&  -\frac{2}{\omega} (
	\eta_{\nu[\sigma}{S}_{\rho]\mu}-\eta_{\mu[\sigma}{S}_{\rho]\nu}
	)\, , \\
	\big\langle s_{\mu\nu}(-\omega)\lambda_{\rho\sigma}(\omega)\big\rangle =& \frac{2}{\omega}\eta_{\mu[\rho}\eta_{\sigma]\nu}\, , \nonumber \\
	\big\langle \lambda^{\mu\nu}(-\omega)z_\rho(\omega)\big\rangle =& -\frac{2}{m\omega}v^{[\mu}\delta^{\nu]}_\rho \, . \nonumber
\end{align}
These propagators should be supplemented with an $i\epsilon$-prescription to be valid for general $\omega$. Standard choices are $\omega\to \omega\pm i\epsilon$ for retarded and advanced propagators, respectively. For the tree-level computations done here this choice is irrelevant since $\omega\neq0$. 
For the metric fluctuation, we use the conventional de Donder gauge propagator,
\begin{align}
\big\langle h_{\mu\nu}(-k)h_{\rho\sigma}(k)\big\rangle = 
\frac{i}{2}\frac{\eta_{\mu \rho}\eta_{\sigma \nu}
		+ \eta_{\mu \sigma}\eta_{\rho \nu} 
		- \eta_{\mu \nu}\eta_{\rho \sigma}}{k^2+i\epsilon}\, ,
\end{align}
where again the Feynman prescription can be dropped for our tree-level computation.

\subsection{Vertices and Compton diagrams}
Simplified graviton self-interactions can be found in ref.~\cite{Kalin:2020mvi} for example, so we turn our attention to constructing vertices involving the worldline fields and gravitons.
For the gravitons it is important to use the Fourier transform
\begin{align}\label{eqn:hfourier}
h_{\mu\nu}(x) = \int\frac{d^4k}{(2\pi)^4} e^{ik\md(b+v\tau + z)}h_{\mu\nu}(k) \, ,
\end{align}
where the linearized solution $x(\tau)$ appears in the exponent. Then, when expanded in powers of $z$, this introduces an infinite number of vertices coupling worldline perturbations to gravitons~\cite{Mogull:2020sak}. The extra $e^{ik\md b}$ factor can be judiciously dropped from the Feynman rules, or alternatively by using the coordinate choice $b^\mu=0$. 
The $\tau$ integrals give us energy-conserving delta functions at each vertex, for example for a vertex with $m$ worldline perturbations and $(n-m)$ gravitons we have,
\begin{align}
\delta( \omega_1+\ldots +\omega_m+k_{m+1} \md v+\ldots +k_n\md v ) \, ,
\end{align}
since all momenta and energies are incoming. As is usually done, we suppress all the momentum and energy conservation delta functions in the Feynman rules below.  

The simplest vertex we have has one graviton sourced by the background worldline trajectory. To construct it, we work in momentum space and keep only terms linear in $h$ and at zeroth order in the worldline fluctuations. Suppressing the coupling $\kappa$ we find
\begin{align}
\begin{tikzpicture}[scale=0.8,baseline={(0, -0.3cm)}]
	\draw[densely dotted] (0,0) -- (2,0);
	\draw[snake it,thick] (1,-1.3) -- (1,0);
\end{tikzpicture}
\ 
=
- \frac{i}{2}m (v \md h\md  v)
- \frac{1}{2} (v\md  h\md  S\md  k) -i\big\langle{\cal L}_{R}\big\rangle\Big|_{h}\, .
\end{align}
We represent the unperturbed worldline with a dotted line, and instead of including free indices in the vertex we used the field $h$ as an off-shell polarization tensor. This notation will become particularly convenient when dealing with the worldline perturbations below. 
We also included the non-minimal interactions through the matrix element $\big\langle {\cal L}_{R}\big\rangle$, which is Fourier transformed to momentum and frequency space and processed in the same way as the contributions from the minimal action.
At second order in the gravitons we have the vertex,
\begin{align}
\begin{tikzpicture}[scale=0.8,baseline={(0, -0.5cm)}]
	\draw[densely dotted] (0,0) -- (2,0);
	\draw[snake it,thick] (0.2,-1.3) -- (1,0);
	\draw[snake it,thick] (1.8,-1.3) -- (1,0);
\end{tikzpicture}
=&
\frac{im}{2}(v\md h_3\md h_4\md  v)
+ \frac{1}{4} (v\md h_3\md h_4\md S\md k_3)
+ \frac{1}{4} (v\md h_3\md S\md h_4\md k_3)
+ \frac{1}{8} {\rm tr}(h_4\md h_3\md S)v\md k_3\nonumber \\
&
+ (3\leftrightarrow 4) -i\big\langle {\cal L}_{R}+  {\cal L}_{R^2}\big\rangle\Big|_{h_3, h_4}\, ,
\end{align}
where now we introduced two placeholder polarizations, $h_3$ and $h_4$. In the Compton amplitude we will replace these with factorized circular polarizations $h_{\mu\nu}\to \varepsilon^\pm_{\mu\nu}=\varepsilon^\pm_\mu \varepsilon^\pm_\nu$.
Finally, we will use one vertex that sources a worldline perturbation as well as a graviton,
\begin{align}
\label{eqn:Uvert}
\begin{tikzpicture}[scale=0.8,baseline={(0, -0.3cm)}]
	\draw[densely dotted] (0,0) -- (1,0);
	\draw[snake it,thick] (1,-1.3) -- (1,0);
	\draw[very thick] (1,0) -- (2,0);
\end{tikzpicture}
=&
\frac{m}{2}(v\md h\md v)(k\md z)
- i (\pi\md  h\md  v)
- \frac{i}{2} (v\md h\md S\md k)(z\md k)
- \frac{1}{2} (v\md h\md s\md k) \nonumber\\
&- \frac{i}{2}(z\md h\md S\md k)\omega+ \frac{1}{2}(v\md h\md v)(v\md \lambda\md  S\md k)
- \frac{1}{2}(v\md h\md v)(v\md s\md k) -i\big\langle {\cal L}_{R}\big\rangle\Big|_{h, W}\, ,
\end{align}
where again we abuse notation slightly by using the worldline fields $\{z^\mu, \lambda^{\mu\nu}, s_{\mu\nu}, \pi_\mu\}=:W$ in the same way as we used $h$ above.
Notice that we represent all possible worldline perturbations with one solid line.

We compute the Compton amplitude to all orders in spin from the worldline action including only linear in Riemann operators $\mathcal{L}_{R}$. We then compare to the all-orders-in-spin amplitude proposed in ref.~\cite{Cangemi:2023bpe} in order to fix $\mathcal{L}_{R^2}$. The Compton amplitude can be computed from the sum of the four diagrams,
\begin{align} \label{WLFeynmanDiagrams}
M(1,2,3^\pm,4^\pm) \  =   \!\!\!\!\!
\begin{tikzpicture}[rotate=180,scale=0.7,baseline={(0, 0.7cm)}]
	\draw[densely dotted] (0,0) -- (4,0);
	\draw[snake it,thick] (1,-2) -- (1,0);
	\draw[snake it,thick] (3,-2) -- (3,0);
	\draw[very thick] (1,0) -- (3,0);
	\node[] at (4.1,0.5) {$p_1{=}mv$};
	\draw[->] (3.3,-1.4) -- (3.3,-0.6);
	\draw[->] (0.7,-1.4) -- (0.7,-0.6);
	\node[] at (0.3,0.5) {$p_2$};
	\draw[->] (2.5,0.4) -- (1.5,0.4);
	\node[] at (2,-0.3) {$\omega$};
	\node[] at (1,-2.3) {$k_3$};
	\node[] at (3,-2.3) {$k_4$};
\end{tikzpicture}
+
\begin{tikzpicture}[rotate=180,scale=0.7,baseline={(0, 0.7cm)}]
	\draw[densely dotted] (0,0) -- (3,0);
	\draw[snake it,thick] (0.5,-2) -- (3/2,0);
	\draw[snake it,thick] (2.5,-2) -- (3/2,0);
	\node[] at (0.5,-2.3) {$k_3$};
	\node[] at (2.5,-2.3) {$k_4$};
\end{tikzpicture}
 + 
\begin{tikzpicture}[rotate=180,scale=0.7,baseline={(0, 0.7cm)}]
	\draw[densely dotted] (0.5,0) -- (3.5,0);
	\draw[snake it,thick] (2,-1) -- (1,-2);
    \draw[snake it,thick] (3,-2) -- (2,-1);
    \draw[snake it,thick] (2,0) -- (2,-1);
	\node[] at (1,-2.3) {$k_3$};
	\node[] at (3,-2.3) {$k_4$};
\end{tikzpicture}
\end{align}
Note that, when expanding the exponential of the action in the path integral, the first diagram in \eqn{WLFeynmanDiagrams} appears with a factor of $\frac{1}{2}$ but it combines with a second identical contribution, hence there is only one massive channel. We take the external momenta to be incoming, and the flow direction of the internal energy $\omega$ is as indicated. This diagram has two delta functions at the vertices $\delta(k_3\md v +\omega)\delta(k_4\md v - \omega)$ which after integrating $\omega$ gives us the same delta function as in the other two diagrams, namely $\delta((k_3+k_4)\md v)$.

See ref.~\cite{Ben-Shahar:2023djm} for example calculations for low orders in spin. Since in this paper we are concerned with all-order-in-spin computations, we will first discuss the general form of a Compton amplitude.

\section{Worldline Compton amplitude to all-orders-in spin}\label{Sect3}

We here compute the Compton amplitude to all orders in spin, using the non-minimal linear-in-$R$ worldline action~\eqref{ppEFT_gfixed_linearR}, or dynamical mass ${\cal M}^2=m^2+{\cal L}_R$. This is done simplest by recycling the notation and pole contributions from \rcites{Cangemi:2022bew,Cangemi:2023ysz,Cangemi:2023bpe}, and then calculating the remaining contact terms using the WQFT Feynman rules. 

\subsection{Classical kinematic variables}

Building upon the notation of \rcites{Cangemi:2022bew,Cangemi:2023ysz,Cangemi:2023bpe}, in addition to the black hole velocity $v^\mu$, we will use natural momentum and spinor combinations 
\begin{align} \label{eq:vars}
& q^\mu := (k_3 + k_4)^\mu  \,, \quad~
q_\perp^\mu := (k_4 - k_3)^\mu \,, \nn \\
& \chi^\mu :=  \langle 3|\sigma^\mu|4]  \,, \quad~~~  X^\mu  := [3|v \sigma^\mu |4] \,,
\end{align} 
where the latter two are complex vectors convenient for encoding the helicity dependence in the opposite- and same-helicity sectors, respectively.
The graviton polarization tensors are factorized into null vectors $\varepsilon_{\mu \nu}^{\pm}=\varepsilon^{\pm}_{\mu} \varepsilon^{\pm}_{\nu}$, and we make the following gauge choices in the two helicity sectors:
\begin{align} \label{eq:polarizations}
(-+)~\text{case}:~~~~ &\varepsilon_3^{\mu-}= \frac{\chi^\mu}{\sqrt{2}[34]}\,,\hskip1.5cm \varepsilon_4^{\mu+}= \frac{\chi^\mu}{\sqrt{2}\langle 34\rangle}\,, \nn \\
(++)~\text{case}:~~~~ &\varepsilon_3^{\mu+}= \frac{X^\mu-2v^\mu[34]}{\sqrt{2} \langle 3|v|4]}\,,~~~
\varepsilon_4^{\mu+}= \frac{X^\mu}{\sqrt{2}\langle 4|v|3]}\,. 
\end{align}
On the second line, the shift by $v^\mu$ is necessary for transversality of $\varepsilon_3^{\mu+}$; however this shift drops out in most natural Lorentz products.  

The corresponding set of non-vanishing Lorentz invariants satisfy
\begin{align} \label{ClassicalLimitScaling1}
 q_\perp^2 & = -q^2  \,, \hskip1.27cm
v \cdot q_\perp =2 \omega \,,  
\qquad \hskip0.51cm
 |q| = \sqrt{-q^2}=  2 \omega \sin \frac{\theta}{2} \,, \nn \\
q_\perp \cdot X& =  - q \cdot X = [34] v\cdot q_\perp   \,, \qquad
v \cdot \chi =\langle 3|v|4]  \,, \qquad  v \cdot X = [34] \,, 
\end{align}
where $\omega$ is the frequency or energy of the graviton planewave (in the rest frame of the black hole), and $\theta$ is the deflection angle for the graviton 3-momentum. 
It is also convenient to introduce the dimensionless variable $\xi:=(v \cdot q_\perp)^2/q^2=4\omega^2/q^2$, known as the optical parameter.

The spin vector $S^\mu$ is related to the ring radius $|a|$ as
\be
S^\mu = m a^\mu 
\ee
where $|a|=\sqrt{-a^2}$. In the classical tree-level Compton amplitude each factor of $a^\mu$ must be accompanied by the graviton frequency or momenta, since their classical scalings ($\hbar\ll1$) are $a^\mu \sim \hbar^{-1}$ and  $\omega \sim \hbar \sim k_i^\mu$. Furthermore, there should not be non-local dependence on $a^\mu$ if we want think of this variable as originating from a local higher-spin QFT~\cite{Cangemi:2023bpe}. It then follows that the classical tree-level Compton amplitude can be written as an entire function in the following dimensionless classical variables\footnote{The $x,y,z$ notation of~\rcite{Cangemi:2023bpe} should not be confused with the similarly named worldline fields.} 
\begin{align} \label{eq:clComptonVars}
x  := a \cdot q_\perp \,, \qquad
y  := a \cdot q \,,   \qquad
z  := |a| \, v \cdot q_\perp \,, 
\end{align}
and the complex variables
\begin{align}\label{eq:clComptonVars2}
w := \frac{a \cdot \chi\;v \cdot q_\perp}{v\cdot\chi} =  \frac{\langle 3|a |4]}{\langle 3|v |4]} v\cdot q_\perp  \,,  \qquad
u :=   \frac{a \cdot X \;v \cdot q_\perp}{v\cdot X} =   \frac{[3|v a |4]}{[34]} v\cdot q_\perp\,,
\end{align}
where $w$ is relevant for $(-+)$ helicity and $u$ for $(++)$ helicity. 

The $z$ variable can be identified with the spheroidicity parameter\footnote{Our $z=2\omega|a|$ has an extra factor of two compared the to conventional choice, see \rcite{Dolan:2008kf}.} used to describe scattering via spheroidal harmonics in the Kerr background. In addition to the spin-dependent classical variables, the Compton amplitude can be a non-trivial function of the optical parameter $\xi$. However, we need to account for the two Gram determinant relations $G(v,q,q_\perp,a,\chi)=0$ and $G(v,q,q_\perp,a,X)=0$, which gives two relations between the optical parameter and the above spin-dependent variables
\be \label{eq:clGramDet}
\xi^{-1}:=\frac{q^2}{(v \cdot q_\perp)^2} =\frac{(w - x)^2 - y^2}{z^2-w^2} =  \frac{(u - y)^2 - x^2}{u^2 - z^2}-1 \,.
\ee
In principle, $\xi$ can be eliminated at the expense of introducing spurious poles. But, in order to capture positive powers of $q^2$ without having spurious poles, we need to allow the Compton amplitude to also be an entire function of the variable $z^2 \xi^{-1}=q^2 |a|^2$. Whenever this variable appears, then one has to simultaneously use \eqn{eq:clGramDet} to remove powers of $w$ and $u$, otherwise the function space becomes over-complete. Thus $q^2 |a|^2$ can at most multiply a single power of $w$ or $u$.

Interestingly, we find that neither the worldline action \eqref{ppEFT_gfixed_linearR} (with ${\cal M}^2=m^2+{\cal L}_R$) nor previous higher-spin QFT considerations~\cite{Cangemi:2023bpe} make use of the $q^2 |a|^2$ variable, thus we will not consider it further in this section.\footnote{One can show that the variable $z^2 \xi^{-1}=q^2 |a|^2$ appears in the Teukolsky computation of \rcite{Bautista:2022wjf}; however, only in the terms proportional to $\alpha$ (introduced to keep track of polygamma contributions). In this paper, we consider the case $\alpha=0$ unless otherwise stated.} 
Note that since $w,u$ are complex, we get two more independent Gram determinant relations if we complex conjugate \eqn{eq:clGramDet}; however, they are automatic if we make complex-conjugation properties of the amplitude manifest. Finally, note that the same-helicity sector and opposite-helicity sectors can be treated almost identically, with the variables $x \leftrightarrow y$ and $u \leftrightarrow w$ swapping roles when moving between the two sectors.

We now give the expected form of Compton amplitudes as obtained from the the worldline before adding $R^2$ operators. For the same-helicity Compton amplitude, we expect the following form\footnote{The coupling $\kappa=4 \pi \sqrt{2 G}$ is suppressed; it can be restored in a Compton amplitude via $M\to\big(\tfrac{\kappa}{2}\big)^2 M$.} 
\begin{align}
M(1,2,3^+,4^+)= m^2\frac{[34]^4 e^{y}}{(v\cdot q_\perp)^2 q^2}-  m^2[34]^4\frac{u^2-z^2}{2(v\cdot q_\perp)^4}\sum_{k=1}^3 u^{3-k} z^{2\lceil \frac{k-1}{2} \rceil} f^{++}_k(x,y,z)\,, \end{align}
where the first term is fixed by the factorization behavior at $q^2=0$ and $v\cdot q_\perp=0$. The remaining terms are local contact terms (even if not manifestly so) and the sum corresponds to three unknown entire functions $f^{++}_k(x,y,z)$ that we will determine. The reason we can factor out $(u^2-z^2)$ is because we expect to find only contact terms that have the property that they vanish for polar scattering\footnote{See \rcite{Bautista:2022wjf} for a proper discussion on the importance, and simplifications, of the polar scattering scenario in the context of solving the Teukolsky equation.} kinematics $u=\pm z$. (This follows from carefully inspecting the linear-in-$R$ worldline action.) The $u$ variable contains a spurious pole that needs to be canceled against the prefactor $[34]^4$, thus $u$ can appear at most up to the fourth power. Likewise, the fourth-order spurious pole in the $v\cdot q_\perp$ variable must be canceled by the same factor appearing inside the $u,z$ variables, thus justifying that these appear as overall polynomials of at least degree four. Finally, we expect only even powers of the spheroidicity $z$ for a conservative process, and the linear-in-$R$ worldline has no dissipative terms.   

For the opposite-helicity Compton amplitude, as obtained from the linear-in-$R$ worldline action, one can write the expected result in two dissimilar forms:
\begin{align}
M(1,2,3^-,4^+)&= m^2\frac{\langle 3|v|4]^4 e^{x-w}}{(v\cdot q_\perp)^2 q^2}-  m^2\langle 3|v|4]^4\frac{w^2-z^2}{2(v\cdot q_\perp)^4}\sum_{k=0}^\infty w^{k} \tilde f^{-+}_k(x,y,z)  \\
&= m^2\frac{\langle 3|v|4]^4 E_{\rm pole}}{(v\cdot q_\perp)^2 q^2}-  m^2\langle 3|v|4]^4\frac{w^2-z^2}{2(v\cdot q_\perp)^4}\sum_{k=1}^3 w^{3-k} z^{2\lceil \frac{k-1}{2} \rceil}  f^{-+}_k(x,y,z)\,.\nn
\end{align}
On both lines the first term corresponds to the known factorization behavior, capturing the $q^2=0$ and $v\cdot q_\perp=0$ poles. The remaining terms have no physical poles (even if not manifestly so), and the sums are over unknown entire functions $\tilde f^{-+}_k$, $f^{-+}_k$ that represent contact terms. On the first line, we use the simple pole term proportional to the well-known exponential $e^{x-w}$~\cite{Guevara:2018wpp,Chung:2018kqs,Chen:2021kxt,Aoude:2022trd,Bautista:2022wjf}; however, this then results in an infinite number of counter terms $\sim w^k \tilde f^{-+}_k$ to cancel out the spurious pole $\langle 3|v|4]^{-1}$ contained in $w$. On the second line, the function $E_{\rm pole}$ represents a spurious-pole-free entire function that matches the physical poles, and we give it in the next section.
The absence of spurious poles $\langle 3|v|4]^{-1}$ then enforces that the infinite sum over contact terms to truncate at third order. Thus we only need to find the three entire functions $f^{-+}_k(x,y,z)$ representing the pole-free contributions. Again, we can factor out $(w^2-z^2)$ since we anticipate finding only contact terms that vanish for polar scattering kinematics $w=\pm z$.

\subsection{Input from higher-spin QFT Compton amplitude}

Before we carry out the worldline computation, it is easier to extract the pole terms $E_{\rm pole}$ from the Compton amplitude of \rcite{Cangemi:2023bpe}, which we now briefly review. The opposite-helicity classical Kerr amplitude, as obtained from the higher-spin (HS) QFT, is
\begin{align}
M_{\rm HS}(1,2,3^-,4^+)  = & \, M_0
\Big\{e^x \cosh z - w \, e^x {\rm sinhc}\, z + \frac{1}{2}(w^2 - z^2)\big( E +2   (x - w) \tilde E \big)\Big\} \nn \\
& ~~~~~ + \text{contact terms}
\end{align} where $E(x,y,z)$ and $\tilde E(x,y,z)$ are entire functions 
\begin{align} \label{eq:clEfns}
E(x,y,z)&:=\frac{e^y - e^x \cosh z + (x - y) e^x \, {\rm sinhc}\,z}{(x - y)^2 - z^2}~+~ (y\to -y) \\
\tilde E(x,y,z)&:=\frac{1}{2y}\frac{e^y - e^x \cosh z + (x - y) e^x \, {\rm sinhc}\,z}{(x - y)^2 - z^2}~+~ (y\to -y) \,.\nn
\end{align} 
with ${\rm sinhc}\,z:=\tfrac{1}{z}\sinh z$, and the classical Schwarzschild amplitude is given by
\be
M_0:= 
\frac{m^2}{q^2\,( v {\cdot} q_\perp)^2} \times \left\{\begin{matrix}
(v \cdot \chi)^4  &~~~~~ (-+)~\text{helicity case}  \\
(v \cdot X)^4 &~~~~~ (++)~\text{helicity case}
\end{matrix}\right. ~~.
\ee
While it is not obvious, the $z$-dependent terms are also contact terms, hence we can further simplify the pole terms by letting $z\to 0$, giving an explicit form for the previously introduced entire function $ E_\text{pole}=E_\text{pole}(x,y,w)$,
\begin{align}
 E_\text{pole}  := &\, \frac{1}{2}
\Big\{e^x(1 - w)+ w^2\frac{y+ x - w}{y} \Big[\frac{e^y-e^x}{(x - y)^2}+\frac{ e^x}{x - y}\Big] \Big\}~+~ (y\to -y) \nn
\\
=&\,1
+ x - w 
+\frac{1}{2} (x - w)^2
+\frac{1}{3!} (x - w)^3
+\frac{1}{4!} \Big((x - w)^4 + w^2 y^2 - w^2 (w - x)^2\Big) \nn \\
&
+\frac{1}{5!} \Big((x - w)^2 (x - 3 w) x^2 - w^2 (w - 3 x) y^2\Big) +\frac{1}{6!}\Big((x - w)^2 (x - 4 w) x^3 \nn \\ 
&~~~~+ w^2 x (5 x - 2 w) y^2 + w^2 y^4\Big) +{\cal O}(a^7)\,.
\end{align}
One can check that the residues on the physical poles are correct. The massive pole corresponds to $v\cdot q_\perp=0$ (such that $w=z=0$) and the amplitude becomes $M_0 e^x$, whereas the massless pole corresponds to $q^2=0$ (such that $\pm y=x-w$) and the amplitude becomes $M_0 e^{\pm y}$. These exponential forms thus agree with the Newman-Janis shift~\cite{Newman:1965tw,Arkani-Hamed:2019ymq}, and the known three-point behavior~\cite{Guevara:2018wpp,Chung:2018kqs}. 

The difference between the manifestly local pole terms and the exponential $e^{x-w}$ can be displayed as
\begin{align}
E_\text{pole}-e^{x-w} & =  - w^2 \big((w - x)^2 - y^2\big)\Big( \frac{1}{4!} - \frac{w-3 x}{5!} + \frac{w^2-4w x+6x^2+y^2}{6!}+{\cal O}(a^3) \Big)
\end{align}
which makes it clear that it vanishes for either $w=0$ or $(w - x)^2 =y^2$, which corresponds to the physical factorization poles discussed above. 

Since in section~\ref{sect:MatchingR^2}, we will work out the worldline ${\cal L}_{R^2}$ terms by comparing to the full higher-spin QFT amplitude, including the explicit contact terms, we will explain them next. In the same-helicity sector, the higher-spin QFT framework~\cite{Johansson:2019dnu, Ochirov:2022nqz,Cangemi:2023bpe} predicts the simple exponential classical amplitude~\cite{Chen:2021kxt}
\begin{align}
M_{\rm HS}(1,2,3^+,4^+)= M_0 e^{y} =  m^2\frac{[34]^4 e^{y}}{(v\cdot q_\perp)^2 q^2}\,,
\end{align}
which agrees with the BHPT results \cite{Bautista:2022wjf} for the Kerr black hole. 

In the opposite-helicity sector, one can write the complete proposed Kerr tree-level amplitude of \rcite{Cangemi:2023bpe} as
\begin{align} \label{Compton_from_HS}
M_{\rm HS}(1,2,3^-,4^+) 
 = & M_0 \Big\{e^x \cosh z - w \, e^x {\rm sinhc}\, z + \frac{1}{2}(w^2 - z^2)\big( E +2   (x - w) \tilde E \big)\Big\} \nn \\
&- \frac{(p \cdot \chi)^4 }{( p {\cdot} q_\perp)^4}\frac{(w^2 - z^2)^2}{2}
\Big( \frac{\partial  \tilde E}{\partial x}+ \eta \frac{\partial  \tilde E}{\partial z} \Big)  + \alpha \, C_\alpha\,,
\end{align}
where the tag $\eta=\pm1$ controls the dissipative terms, and $\alpha$ is an auxiliary parameter that tags certain contributions related to polygamma functions, which we associate with an unknown $C_\alpha$ contribution. These tags were introduced in~\rcite{Bautista:2022wjf}, and the amplitude \eqref{Compton_from_HS} is in agreement with the finite spin-mutipole results of~\rcite{Bautista:2022wjf,privateBGKV, Bautista:2023sdf} for $\alpha=0$.
In superextremal limit $|a|\gg Gm$ the dissipative terms become real (non-imaginary) by analytic continuation and proportional to the spin magnitude $|a|$~\rcite{Bautista:2022wjf}, or in our preferred variables, proportional to $z$. Likewise, the $\alpha$-dependent terms appear to be proportional to $z^2$ up to order $S^8$ where they are currently known~\cite{Bautista:2022wjf,Bautista:2023sdf}. In this paper, we assume that the tree amplitude is characterized by $\alpha=0$, so that we can focus on reproducing the all-orders-in spin prediction of ref.~\cite{Cangemi:2023bpe}.  

For reference, we display the first few multipole orders of the entire functions discussed above
\begin{align}
E(x,y,z)&=1 + \frac{2x}{3}+ \frac{1}{12}(3 x^2 + y^2 + z^2) + 
 \frac{x}{30} (2 x^2 + y^2 + 2 z^2) + {\cal O}(a^4)\,,  \nn \\
 \tilde E(x,y,z)&=
\frac{1}{6} +  \frac{x}{12} +\frac{1}{120} (3 x^2 + y^2 + z^2) + 
 \frac{x}{360} (2 x^2 + y^2 + 2 z^2)  + {\cal O}(a^4)\,, \nn \\
{\cal E}(x,y,z)&=\frac{\partial  \tilde E}{\partial x}=
\frac{1}{12} + \frac{x}{20}  +\frac{1}{360}  (6 x^2 + y^2 + 2 z^2) + \frac{x}{2520}(10 x^2 + 3 y^2 + 10  z^2) +{\cal O}(a^4)\,, \nn \\
\tilde {\cal E}(x,y,z)&=\frac{\partial  \tilde E}{\partial z}=
\frac{z}{60} + \frac{x z}{90} + \frac{z}{2520}(10 x^2 + y^2 + 2 z^2)  +{\cal O}(a^4)\,.
\end{align}
Similarly, the full opposite-helicity amplitude \eqref{Compton_from_HS} has the expansion
\begin{align} \label{Ampl_HS_series_exp}
M_{\rm HS}(1,2,3^-,4^+)\big|^{\alpha=0}_{\eta=0} = & M_0 \Big\{1 + x - w
+\frac{1}{2} (x - w)^2
+\frac{1}{3!}(x - w)^3
+\frac{1}{4!} (x - w)^4 \nn\\
&+\frac{1}{5!}\Big((x - w)^2 x (3 w^2 - 3 w x + x^2) - w^3 y^2 - 
    z^2 w ((x - w)^2 - y^2)\Big) \nn\\
&+\frac{1}{6!} \Big((x-w)^2 x^2 (6 w^2 - 4 w x + x^2) + w^3 y^2 (w - 4 x) \nn\\ &~~~~~~+ z^2 ((x - w)^2 - y^2) (2 w (w - 2 x) - z^2)\Big)\Big\} + {\cal O}(a^7)\,,
\end{align}
and we give the dissipate terms separately 
\be
M_{\rm HS}(1,2,3^-,4^+)\big|^{\alpha=0}_{\eta} = M_0  \eta  z  (w^2 - z^2) ((x - w)^2 - y^2) \Big(\frac{1}{5!} + \frac{4 x}{6!} \Big)+ {\cal O}(a^7)\,.
\ee
As seen in \eqn{Ampl_HS_series_exp}, the first four spin orders match the expected exponential pattern, and thereafter the amplitude becomes slightly more complicated. However, the numerical coefficients are small integers divided by the expected factorial denominators. Indeed, one can confirm by explicitly expanding out the first 100 orders that the numerical coefficients are always small integers over the appropriate factorial of that spin order. By ``small'' we mean that they grow at worst as $\sim 2^n$ in the spin order $n$, which is much slower than the factorial $n!$ in the denominator.   

\subsection{Opposite-helicity Compton amplitude, with linear-in-$R$ terms}
After computing the Compton amplitude from the linear-in-Riemann worldline action \eqref{ppEFT_gfixed_linearR} up to order $S^{16}$, we plug in the polarization tensors corresponding to the opposite-helicity case \eqref{eq:polarizations}. From the obtained expressions we can extrapolate the pattern to all orders, which sum to the following entire functions:
\begin{align} \label{WL_opposite_entire_fn}
f_3^{-+}:=&\frac{2 e^x x + 2 x + x^2 - y^2 - 
 (4 x + x^2 - y^2) e^{\tfrac{x}{2}} \cosh\tfrac{y}{2}}{(x^2 - y^2)^2}\,,
\nn \\
f_2^{-+} := &\frac{8 e^x x - (2 + 2 e^x + x) (8 + x^2 - y^2)-\tfrac{1}{4}(x^2 - y^2)^2}{(x^2 - y^2)^3} \nn \\ &  
+e^{\tfrac{x}{2}} \frac{(32 + x^3 - x y^2) \cosh\tfrac{y}{2} - y (16 + x^2 - y^2)  \sinh\tfrac{y}{2} }{(x^2 - y^2)^3}\,,  \\
f_1^{-+}:=&
\frac{8 x (2 x + x^2 - y^2) e^{\tfrac{x}{2}}  \cosh\tfrac{y}{2}}{(x^2 - y^2)^3} 
+ \frac{4  y (4 x - x^2 + y^2) e^{\tfrac{x}{2}}\sinh\tfrac{y}{2}}{(x^2 - y^2)^3}+ \frac{e^{y}}{2 y^2 (x - y)^2} 
+ \frac{e^{-y}}{2 y^2 (x + y)^2}  \nn \\ & 
- \frac{x^2 + 3 y^2}{y^2 (x^2 - y^2)^2} 
- \frac{3}{2}\frac{x^2 - y^2 + 4 x}{(x^2 - y^2)^2} 
- \frac{16 x^2 e^{x}}{(x^2 - y^2)^3 }
+ \frac{2 e^{x}  (x + 1)}{(x^2 - y^2)^2}
\,,\nn \\
f_0^{-+}:=&
\frac{e^{x} (x+4) +5 x+ 4  + \frac{3}{2}  (x^2 - y^2) - 
   \big(8 + 6 x + \frac{1}{2}(x^2 - y^2) \big) e^{\tfrac{x}{2}} \cosh\tfrac{y}{2} 
 + 4 y e^{\tfrac{x}{2}} \sinh\tfrac{y}{2}}{(x^2 -y^2)}\,. \nn
\end{align}
Thus the final opposite-helicity Compton amplitude as obtained from the worldline action \eqref{ppEFT_gfixed_linearR} is given by
\begin{align} \label{AmpWL_opposite_h}
M(1,2,3^-,4^+) =&  \frac{M_0}{2}
\Big\{e^x(1 - w)+ w^2\frac{y+ x - w}{y} \Big[\frac{e^y-e^x}{(x - y)^2}+\frac{ e^x}{x - y}\Big] ~+~ (y\to -y) \Big\} \nn \\
&- m^2\frac{ (p\cdot \chi)^4 }{(p\cdot q_\perp)^4} \frac{w^2 - z^2}{2}\Big(w^2 f_1^{-+}+ w z^2 f_2^{-+}+z^2 f_3^{-+}+ \beta (w^2-z^2) f_0^{-+} \Big) 
\end{align}
where we have introduced a toggle parameter $\beta$ for the two alternative prescriptions \eqref{P1_prescription} and \eqref{P2_prescription} for the worldline action~\eqref{ppEFT_gfixed_linearR},
\be
\beta = \left\{\begin{matrix}
1 & ~~~\hat{p}^\mu \to p^\mu/|p|~\text{prescription, i.e.~}{{\cal L}_R} \\
0 & ~~~\hat{p}^\mu \to p^\mu/m~\text{ prescription, i.e.~}{{\cal L}_R}'  
\end{matrix}\right.
\ee
We note a curious feature: the worldline action gives entire functions that typically depend on hyperbolic functions with half-value arguments $x/2,y/2$, whereas the QFT results~\cite{Cangemi:2023bpe} gave hyperbolic functions with whole-value arguments $x,y$. This mismatch of the argument is both surprising and somewhat concerning since the worldline action \eqref{ppEFT_gfixed_linearR} incorporates linear-in-Riemann interactions that are well-established in the literature. This mismatch between the expected answer and the worldline gets even worse in the same-helicity sector. 

Since the all-spin-orders formula \eqref{AmpWL_opposite_h} maybe difficut to digest for the reader, we can series expand it for the case $\beta=1$, which gives
\begin{align}
&M(1,2,3^-,4^+)\big|_{\beta=1}= M_0 \Big[ 1
+x - w
+\frac{1}{2} (x - w)^2 
+\frac{1}{3!} (x - w)^3 
+\frac{1}{4!} (x - w)^4  \nn \\ &
+\frac{1}{2^3 6!} \Big(3  x (x - w)^2 (37 w^2 - 48 w x + 16 x^2) 
+ 3 w^2 y^2 (11 x - 16 w) + 5 z^2 (3 x - 10 w) \\ & ~~~~~~~~~~\times ((w - x)^2 - y^2)\Big) +\frac{1}{2^5 6!}\Big(
 w^2 y^2 (29 w^2 - 122 w x + 68 x^2 + 3 y^2) +x^2 (x - w)^2  \nn \\ &
~~~~~~~~~~\times (121 w^2 - 128 w x + 32 x^2)  
 + 
 5 z^2 (5 x^2 + y^2 - 14 w x) ((w - x)^2 - y^2)\Big) +{\cal O}(a^7) \Big]\,. \nn 
\end{align}
On the first line, we see that the worldline amplitude is an exponential up to $S^4$, but at higher orders the terms exhibit no structure and numerical coefficients are rather large fractions. We can interpret this as an indication that the linear-in-$R$ worldline action does a poor job at giving reasonable $S^{\ge 5}$ results. For example, compare to the corresponding amplitude obtained from higher-spin theory~\eqref{Ampl_HS_series_exp} which has a much cleaner expansion.  Note that the alternative prescription $\beta=0$ does even worse, as the exponential pattern only holds up to $S^3$, and every order thereafter is more complicated, e.g.
\begin{align}
M(1,2,3^-,4^+)\big|_{\beta=0}= M_0 \Big[ & 1
+x - w
+\frac{1}{2} (x - w)^2 
+\frac{1}{3!} (x - w)^3 \nn \\
&+\frac{1}{4!} \Big((x - w)^4-\frac{3}{2} \big((w - x)^2 - y^2\big) (w^2 - z^2)\Big) +{\cal O}(a^5) \Big]\,, \nn 
\end{align}
where we spare the reader from the complicated higher-order terms.

\subsection{Same-helicity Compton amplitude, with linear-in-$R$ terms}
We now consider the same-helicity Compton amplitude as obtained from the linear-in-Riemann worldline action \eqref{ppEFT_gfixed_linearR}. We evaluate the previously computed amplitude (up to order $S^{16}$) for the case that both graviton polarization tensors have positive helicity. We then extrapolate the pattern to all orders, which sum to the following entire functions:
\begin{align} \label{f++definitions}
f_0^{++}:=&  f_0^{-+}\Big|_{x\leftrightarrow y} \,, \nn \\
f_1^{++}:=& - f_3^{-+}\Big|_{x\leftrightarrow y} \,,\nn \\
f_2^{++}:=&  \Big(f_2^{-+} -\frac{1}{4} f_3^{-+}+ \frac{1}{2} f_0^{-+}\Big)\Big|_{x\leftrightarrow y}\,,  \\
f_3^{++} :=& \frac{8  x (2 y + y^2 - x^2) e^{\tfrac{y}{2}}\sinh\tfrac{x}{2}}{(y^2 - x^2)^3} + 
\frac{4 (4 x^2 + y (x^2 - y^2)) e^{\tfrac{y}{2}}  \cosh\tfrac{x}{2} }{(y^2 - x^2)^3}\nn \\ &
 + \frac{x^2 + x^2 y + 5 y^3}{ y^2 (x^2 - y^2)^2} 
 - \frac{3}{2}\frac{x^2 - y^2 - 2}{(x^2 - y^2)^2} 
 + \frac{e^y ( 17 y^4 - x^4) }{y^2 (x^2 - y^2)^3}
 + \frac{2 e^y (y + 7)}{(x^2 - y^2)^2}\,, \nn 
\end{align}
where some of the entire functions are recycled from \eqn{WL_opposite_entire_fn}. Thus, the final equal-helicity Compton amplitude, as obtained from the worldline action \eqref{ppEFT_gfixed_linearR}, is given by
\begin{align} \label{AmpWL_same_h}
M(1,2,3^+,4^+) 
 =M_0 e^{y} - m^2\frac{[34]^4}{(v\cdot q_\perp)^4}  \frac{u^2 - z^2}{2}  \Big(u^2 f_1^{++}+ u z^2 f_2^{++} +z^2 f_3 + \beta (u^2 - z^2) f_0^{++}\Big) .
\end{align}
Clearly, only the first exponential term is what we expect to describe a Kerr black hole, thus all the other terms seem to be unwanted ``garbage''. These need to be subtracted out using $R^2$ operators. 

To make the amplitude \eqref{AmpWL_same_h} more explicit, we series expand it in spin multipoles for the prescription $\beta=1$, 
\begin{align}
M(1,2,3^+,4^+)\big|_{\beta=1} = &\, M_0\Big[1+ y+ 
\frac{y^2}{2}+ 
\frac{y^3}{3!}+ 
\frac{y^4}{4!} \\ 
& ~~~~ +\frac{1}{2^3 6!} \Big(48 y^5 + (x^2 - y^2 - z^2 + 2 u y) (40 u z^2 - 15 u^2 y + 3 y z^2)\Big)
\nn \\ 
& ~~~~ +\frac{1}{2^5 6!} \Big(32 y^6 - (x^2 - y^2 - z^2 + 2 u y) \big(5 u^2 (x^2 + 5 y^2) \nn  \\
  & \hskip5cm  + z^2(5 x^2 - 50 u y - 7 y^2)\big)\Big) \Big] +{\cal O} (a^7)\,, \nn
\end{align}
as can be seen the same large fractions encountered in \eqn{AmpWL_opposite_h} appear beyond $S^4$, which suggests that the similar unwanted contributions pollute both the same and opposite helicity sectors of the worldline action. Again, choosing the alternative prescription $\beta=0$ makes in general every multipole more complex, and the exponential pattern $e^y$ breaks already at order $S^4$,
\begin{align}
M(1,2,3^+,4^+)\big|_{\beta=0} = &\, M_0\Big[1+ y+ 
\frac{y^2}{2}+ 
\frac{y^3}{3!} + \frac{1}{48} \big(2 y^4 - 3 (u^2 - z^2) (x^2 - y^2 - z^2+ 2 u y)\big)\Big] \nn \\ &~~~~+{\cal O} (a^5)\,.
\end{align}
Note that the unwanted terms are always proportional to $(x^2 - y^2 - z^2 + 2 u y)\propto (u^2-z^2)$, which vanish for polar-scattering kinematics.

\section{Matching $R^2$ operators: from QFT to worldline}
\label{sect:MatchingR^2}

We will now work out the $R^2$ operators needed for the dynamical mass function of the worldline action~\eqref{eq:dynamicalMass}, in order for the worldline Compton amplitude to reproduce the results coming from the higher-spin QFT framework of~\rcite{Cangemi:2023bpe}.
In effect, we present the worldline action that corresponds the far-zone part of the Teukolsky solution. Generically, this far-zone contribution should be corrected by loops; however, we expect that terms in the amplitude that do not explicitly depend on the Casimir $S^2$ (or $|S|$) are protected from such corrections. This is because, by dimensional analysis, for an $L$-loop contribution from a particular operator to mix with the tree-level amplitude its Wilson coefficient in a standard worldline action must come with a factor of $|S|^L/G^L$.

We first consider a complete classification of independent $R^2$ operators, to all orders in spin, and then we give the explicit results.

\subsection{Basis of $R^2$ operators, conservative sector}

The Riemann-square operators are most conveniently built out of the following three curvature combinations:
\begin{align}
&E^-_{\mu\nu}E^+_{\rho\sigma}=\frac{1}{4}(E_{\mu\nu}E_{\rho\sigma}+B_{\mu\nu}B_{\rho\sigma})+\frac{i}{4}(E_{\mu\nu}B_{\rho\sigma}-B_{\mu\nu}E_{\rho\sigma}) \,,\nn \\
&E^+_{\mu\nu}E^+_{\rho\sigma} = \frac{1}{4}(E_{\mu\nu}E_{\rho\sigma}-B_{\mu\nu}B_{\rho\sigma})+\frac{i}{4}(E_{\mu\nu}B_{\rho\sigma}+B_{\mu\nu}E_{\rho\sigma})\,,  \nn \\
&E^-_{\mu\nu}E^-_{\rho\sigma} = \frac{1}{4}(E_{\mu\nu}E_{\rho\sigma}-B_{\mu\nu}B_{\rho\sigma})-\frac{i}{4}(E_{\mu\nu}B_{\rho\sigma}+B_{\mu\nu}E_{\rho\sigma})\,,
\end{align}
where we recall that the electric and magnetic curvature tensors are defined as
\begin{align}
E^{\pm}_{\mu \nu}&:= \frac{1}{2}(E_{\mu \nu} \pm i B_{\mu \nu}) \,, \nn \\
E_{\mu \nu} &:= \hat p^\rho \hat p^\sigma R_{\mu \rho \nu \sigma}  \,,\nn \\
B_{\mu \nu} &:= \frac{1}{2}\hat p^\rho \hat p^\sigma {\epsilon_{\nu\sigma}}^{ \kappa \lambda}  R_{\mu \rho \kappa \lambda}\,.
\end{align}
The chiral $E^{\pm}_{\mu \nu}$ tensors are more convenient to work with than the electric and magnetic curvatures, as they are in one-to-one correspondence with the helicity sectors of the Compton amplitude. Note, when dealing with quadratic in curvature operators there is no difference between using $\hat{p} = p^\mu/m$ or $\hat{p}^\mu = p^\mu/\sqrt{p^2}$ for the Compton amplitude. However, we assume that the default prescription is that $\hat{p}^\mu = p^\mu/\sqrt{p^2}$, unless otherwise stated.  

Starting with the above quadratic curvature tensors, we then need to include dependence on the remaining worldline fields and derivatives, which we take to be $\{p_\mu,S^\mu,\nabla_\mu, g^{\mu \nu}\}$. Based on the observed structure of our Compton amplitudes, there is a natural split of operators into two classes: those that vanish for polar scattering and those that do not. 
Under the polar scattering kinematics, the 3-momenta of the incoming graviton planewave is parallel to the spin 3-vector. The curvature tensor of such a planewave satisfies
\be
S^\nu E_{\mu \nu}^{\pm}=E_{\mu S}^{\pm}= 0\,.
\ee
Note that ${\cal L}_{R}$ is written in terms of such operators $E^{\pm}_{S\mu}$, suggesting that they are well-suited for describing a Kerr black hole at higher-orders. 
For $R^2$ operators, crossing symmetry implies that the incoming planewave contributes to both curvature tensors and we must build the {\it polar-vanishing} (pv) operators out of two factors of $E_{S \mu}^{\pm}$, and the {\it polar-contributing} (pc) operators can have at most one $E_{S \mu}^{\pm}$ factor. \\[5pt]
\noindent
{\bf Operators that vanish for polar scattering:} \\
We now introduce a basis of conservative polar-vanishing (con-pv) quadratic-in-Riemann operators,\footnote{We pulled out an overall factor of $2^5$ in \eqn{ActionPolarS} to remedy a proliferation of such factors elsewhere. In the Compton amplitude such powers of two cancel out, see \app{appendix_action_to_onshell}.}
\be \label{ActionPolarS}
{\cal L}_{R^2}^\text{con-pv}=\frac{2^5}{m^2}\big(E^-_{S\mu} O_{-+}^{\mu \nu} E^+_{S\nu}+E^+_{S\mu} O_{++}^{\mu \nu} E^+_{S\nu}\big) ~+~{\rm h.c.}\,,
\ee
where every curvature tensor is contracted with one spin vector, and the remaining Lorentz indices have slightly more complicated structure. 
The differential operators $O_{\pm\pm}^{\mu \nu}$ account for the last contraction, and they take the form
\begin{align}\label{Ooperators}
O_{\pm\pm}^{\mu \nu} &:=  \Big(S^\mu S^\nu
- \frac{1}{2} g^{\mu\nu} S^2\Big) {\cal F}^{\pm \pm}_1   - \frac{i}{m} S^2 {\cal F}^{\pm \pm}_2 \Big(S^\nu \nabla^\mu- g^{\mu\nu}  S\cdot \nabla\Big)-\frac{1}{2}g^{\mu\nu} S^2 {\cal F}^{\pm \pm}_3 \ .
\end{align}
In turn, these differential operators contain the entire functions
\be
{\cal F}^{\pm \pm}_k={\cal F}^{\pm \pm}_k(d,b,\vartheta)\,,
\ee
that depend on the three spin-dependent first-order differential operators
\def\zop{\vartheta}
\def\bop{b}
\def\dop{d}
\be \label{differentialOps}
\dop := -\frac{i}{m} S \cdot \nabla \,,~~~~~~\bop :=  \overset{\leftarrow}{\nabla} \cdot S  \frac{i}{m} \,,~~~~~~   \zop := -\frac{2i}{m} |S|\, \hat{p}\cdot \nabla\,.~~
\ee

Because of the explicit hermitian conjugate in \eqn{ActionPolarS} and the symmetry of its second term, there is some unwanted redundancy in the Lagrangian. This redundancy can be removed by expanding the operators in the Lagrangian to lowest order in the coupling and imposing that $O_{-+}^{\mu \nu}$ is manifestly hermitian, and that $O_{++}^{\mu \nu}$ is manifestly symmetric (under operator transpose), this gives the following constraints on the ${\cal F}^{\pm\pm}_k$ functions:
\begin{align} \label{F_Hermiticity}
{\cal F}^{-+}_k(\dop,\bop,\zop)& = \Big({\cal F}^{-+}_k(-\bop,-\dop,-\zop)\Big)^* \,, \nn \\
{\cal F}^{++}_k(\dop,\bop,\zop) & = {\cal F}^{++}_k(-\bop,-\dop,-\zop) \,,
\end{align}
Here we used the differential operator transpose $(\nabla)^t= \overset{\leftarrow}{\nabla}$, and use integration by parts for $\hat p\cdot \nabla$, the latter is valid only in the tree-level contribution of the contact terms to the Compton amplitude, because $\hat p\cdot \nabla\approx \partial/\partial\tau\approx -\overset{\leftarrow}{\nabla}\cdot \hat p$ by use of the equations of motion. Note that any contributions to ${\cal F}^{\pm\pm}_k$ that do not respect the above constraints will correspond to higher-order in $\kappa$, and can thus be relegated to cubic-in-Riemann operators. 

Hermiticity of the Lagrangian is equivalent to enforcing $CPT$ symmetry. Furthermore, in order to describe a Kerr black hole, we wish to enforce parity symmetry, and we can assume that $C$ acts trivially on the fields of the worldline action. Taking into account that helicities of the curvature tensors flip under parity transformations, $P E^+_{ij}=E^-_{ij}$, and that $d,b$ are parity-odd, we can deduce that this implies the following parity constraints: 
\begin{align} \label{F_Psymmetry}
P\text{-even}:~~~~~~  {\cal F}^{\pm\pm}_k(\dop,\bop,\zop) &= \Big({\cal F}^{\pm\pm}_k(-\dop,-\bop,\zop) \Big)^*\,, 
\end{align}
where Hermitian conjugate for the variables are $b^\dagger=d$, $d^\dagger=b$ and $\zop^\dagger= \overset{\leftarrow}{\nabla}{\cdot} \hat p \frac{2i}{m} \approx \zop $ where the last equality only holds at leading order in the perturbative expansion.
Finally, we want to impose time-reversal symmetry on the conservative Lagrangian. It takes the form
\begin{align} \label{F_Tsymmetry}
T\text{-even}:~~~~~~  {\cal F}^{\pm\pm}_k(\dop,\bop,\zop) &= \Big({\cal F}^{\pm\pm}_k(-\dop,-\bop,-\zop) \Big)^*\,,  
\end{align}
Comparing \eqnss{F_Hermiticity}{F_Psymmetry}{F_Tsymmetry}, we see that they can be equivalently phrased as the following conditions:
\begin{align} \label{final_F_constraints}
{\cal F}^{-+}_k(\dop,\bop,\zop)& = {\cal F}^{-+}_k(\bop,\dop,\pm\zop)\,,~~~~~~~~ \text{ coefficients in } {\cal F}^{-+}_k \text{ real}\,, \nn \\
{\cal F}^{++}_k(\dop,\bop,\zop) & = {\cal F}^{++}_k(-\bop,-\dop,\pm\zop)\,,~~~~ \text{ coefficients in } {\cal F}^{++}_k \text{ real}\,, 
\end{align}
Thus all the functions are even in $\zop$, and their coefficients are real, and the ${\cal F}^{-+}_k$ functions are symmetric under exchange $\dop \leftrightarrow \bop$, while the ${\cal F}^{-+}_k$ functions are symmetric under exchange $\dop \leftrightarrow  -\bop$.

We can now count how many free parameters the entire functions contain, which is easiest done using a generating function that tag the spin-multipole order. 
A three-variable {\it real entire function} has the freedom corresponding to the triangular numbers, giving the generating function
\be
\sum_{n=0}^{\infty} \frac{1}{2} (n+1) (n+2)t^n =\frac{1}{(1 - t)^3} ~~~~\stackrel{P,T}{\longrightarrow} ~~~~ \frac{1}{(1-t)^3(1+t)^2} 
\ee
where on the right we imposed the constraints due to $P$ and $T$ symmetries. This is equivalent to requiring that the real entire function is even in two of the variables.\footnote{In general, a real entire function of $n=p+q$ variables, that is even in $q$ variables, has the freedom corresponding to the generating function $(1-t)^{-n}(1+t)^{-q}= (1-t)^{-p}(1-t^2)^{-q}$.}

Finally, we add up the freedom for the entire functions according to which spin-multipole order they start contributing: ${\cal F}^{\pm\pm}_1$ and ${\cal F}^{\pm\pm}_2$ start contributing at $S^4$ due to their prefactors, and ${\cal F}^{\pm\pm}_3$ start at $S^5$, thus we get the following number of free Wilson coefficients
\be
2 \times \frac{2t^4+t^5}{(1-t)^3(1+t)^2} = 2\big(2 t^4 + 3 t^5 + 7 t^6 + 9 t^7 + 15 t^8 + 18 t^9 + \ldots\big)\,,
\ee
where the overall factor of 2 takes into account that the same-helicity sector ${\cal F}^{++}_k$, contains equally many independent operators as the opposite-helicity sector, ${\cal F}^{-+}_k$. 
Hence in both sectors, at order $S^4$ we have to determine $2\times 2$ free Wilson coefficients, at $S^5$ we have $2\times 3$ coefficients, etc. 

We can now evaluate the Compton matrix elements of the operators, they are
\begin{align}\label{O_to_F_ops}
\frac{2^5}{m^2}\Big\langle E^-_{S\mu} O_{-+}^{\mu \nu} E^+_{S\nu} \Big\rangle &= \frac{(v \cdot \chi)^4}{2(v \cdot q_\perp)^4}(w^2-z^2)(w^2 {\cal F}^{-+}_1+w z^2 {\cal F}^{-+}_2+ z^2 {\cal F}^{-+}_3)\,, \nn \\
\frac{2^5}{m^2}\Big\langle E^+_{S\mu} O_{++}^{\mu \nu} E^+_{S\nu} \Big\rangle & =\frac{(v \cdot X)^4}{2(v \cdot q_\perp)^4} (u^2-z^2)(u^2 {\cal F}^{++}_1+ u z^2 {\cal F}^{++}_2+ z^2 {\cal F}^{++}_3)\,, \nn \\
\frac{2^5}{m^2}\Big\langle E^-_{S\mu} O_{--}^{\mu \nu} E^-_{S\nu} \Big\rangle &=  \frac{(v \cdot \bar X)^4}{2(v \cdot q_\perp)^4} (\bar u^2-z^2)(\bar u^2 {\cal F}^{--}_1+  \bar u z^2  {\cal F}^{--}_2+z^2 {\cal F}^{--}_3)\,,
\end{align}
where we used the complex conjugate $\bar u= u^*$, and the ${\cal F}$'s are now functions of the kinematic variables ${\cal F}^{\pm\pm}_k={\cal F}^{\pm\pm}_k(\frac{x+y}{2},\frac{x-y}{2},z)$ according to the on-shell evaluation of the differential operators,
\begin{equation}\label{on_shell_diff_ops}
    d \to \frac{x+y}{2} \ , \ \ \ \ b\to \frac{x-y}{2} \ , \ \ \ \ \vartheta\to z \ .
\end{equation}
For further details on the on-shell evaluation of the action, see appendix \ref{appendix_action_to_onshell}.
In terms of $x,y,z$ the functions satisfy
\begin{align} \label{xyzSymetries}
&{\cal F}^{-+}_k~~~:~~~\text{even in both}~~y,z \nn \\
&{\cal F}^{++}_k~~~:~~~\text{even in both}~~x,z \nn \\
&{\cal F}^{--}_k = {\cal F}^{++}_k\Big|_{y\to -y}\,.
\end{align} \\[5pt]
\noindent
{\bf Operators that contribute to polar scattering:} \\
 We will now complete the discussion of conservative $R^2$ by introducing the remaining polar-contributing operators. Note that we find such operators to be absent in the worldline action when matching the Kerr Compton amplitude.
In general, a complete basis of $R^2$ operators also requires the introduction of 
\be \label{LR2_conservative_pc}
{\cal L}_{R^2}^\text{con-pc} = \frac{2^5}{m^2}g^{\rho \sigma} \Big(E^{-}_{\mu \rho}  \widetilde O_{-+}^{\mu \nu}  E^{+}_{\nu \sigma}+E^{+}_{\mu \rho} \widetilde O_{++}^{\mu \nu}  E^{+}_{\nu \sigma}\Big)~+~{\rm h.c.}\,,
\ee
where
\be \label{OtidleOperator}
\widetilde O_{\pm \pm}^{\mu \nu}:=   \frac{1}{2}g^{\mu\nu} S^4 {\cal G}_1^{\pm \pm}  + \frac{i}{m} S^4 {\cal G}_2^{\pm \pm} (S^\nu \nabla^\mu- g^{\mu\nu}  S\cdot \nabla)\,,
\ee
where the ${\cal G}_k^{\pm\pm}$ are entire functions that depend on four differential operators
\be \label{iotaDef}
{\cal G}_k^{\pm\pm}={\cal G}_k^{\pm\pm}(d,b,\vartheta,\iota)\,,~~~~~~~   \iota:=  \frac{2\overset{\leftarrow}{\nabla}_\mu S^2  \nabla^\mu}{m^2}\,,~~~~~~~\iota^\dagger=\iota\,,
\ee
where $d,b,\vartheta$ were defined \eqn{differentialOps} and the fourth differential operator $\iota$ is new. 

We can now evaluate the Compton matrix elements of the operators, they are
\begin{align}
\frac{2^5}{m^2}\Big\langle g^{\rho\sigma} E^-_{\mu\rho} \tilde O_{-+}^{\mu \nu} E^+_{\nu\sigma} \Big\rangle &= \frac{(v \cdot \chi)^4}{2(v \cdot q_\perp)^4}(z^4 {\cal G}^{-+}_1+ w z^4 {\cal G}^{-+}_2)\,, \nn \\
\frac{2^5}{m^2}\Big\langle g^{\rho\sigma} E^+_{\mu\rho} \tilde O_{++}^{\mu \nu} E^+_{\nu\sigma} \Big\rangle & = \frac{(v \cdot X)^4}{2(v \cdot q_\perp)^4} (z^4 {\cal G}^{++}_1+ u z^4 {\cal G}^{++}_2)\,, \nn \\
\frac{2^5}{m^2}\Big\langle g^{\rho\sigma} E^-_{\mu\rho} \tilde O_{--}^{\mu \nu} E^-_{\nu\sigma} \Big\rangle &= \frac{(v \cdot \bar X)^4}{2(v \cdot q_\perp)^4} ( z^4 {\cal G}^{--}_1+ \bar u z^4  {\cal G}^{--}_2)\,,
\end{align}
where the ${\cal G}^{\pm\pm}_k$s are now functions of the kinematic variables ${\cal G}^{\pm\pm}_k={\cal G}^{\pm\pm}_k(\frac{x+y}{2},\frac{x-y}{2},z,q^2|a|^2)$. The ${\cal G}^{\pm\pm}_k$s satisfy the exact same hermiticity properties and $P$- and $T$-symmetry as the ${\cal F}^{\pm\pm}_k$s; see \eqn{F_Hermiticity} through \eqref{final_F_constraints}, as well as \eqn{xyzSymetries}. The ${\cal G}^{\pm\pm}_k$ functions depend on four variables while the ${\cal F}^{\pm\pm}_k$ functions on three due to the Gram determinant relation \eqref{eq:clGramDet} that can be used for the polar-vanishing operators such that $q^2|a|^2$ can be totally eliminated. Since the polar-contributing operators give matrix elements with at most single powers of $w$ and $u$, the Gram determinant relation cannot be used while maintaining locality, as explained previously. 

Thus, after adding the ${\cal L}_{R^2}^\text{con-pv}$ and ${\cal L}_{R^2}^\text{con-pc}$ together, the total number of independent conservative $R^2$ operators is given by the generating series\footnote{Note that we can in principle allow the Wilson coefficients of the independent operators to functions of the dimensionless quantity $|a|/(Gm)$, and thus the classification does not count powers of this number as independent operators.}
\be \label{OperatorCount_Conservative}
2\Big(\frac{2t^4 + t^5}{(1 - t)^3 (1 + t)^2}+ \frac{t^4 + t^5}{(1 - t)^4 (1 + t)^3}\Big) = 2\big(3 t^4 + 5 t^5 + 12 t^6 + 17 t^7 + 29 t^8 + 38 t^9 + 56 t^{10} +\ldots \big)
\ee
where we recall that the $(1-t)^n$ denominator describes an unconstrained real entire function of $n$ variables, the $(1+t)^{n-1}$ denominator imposes the restriction that it is even in $n-1$ of those variables, and the numerators tell us how many such entire functions we have and the spin-multipole order of their prefactors. The overall factor of 2 is again because the same-helicity and opposite-helicity sectors contribute equally to the count. The count in \eqn{OperatorCount_Conservative} agrees with that of \rcite{Haddad:2023ylx}.

\subsection{Final results for conservative $R^2$ operators}

We now give the same-helicity differential functions, given in terms of four entire functions $\frak{D}_{0,1,2,3}$. For helicities $(++)$, we have
\begin{align}
{\cal F}^{++}_1(\dop,\bop,\zop)& =  \frak{D}_1(\dop,\bop)+ \beta \frak{D}_0(\dop,\bop)\,,\hskip 1.5cm {\cal G}_1^{++}(\dop,\bop,\zop,\iota) =0\,, \nn \\ 
{\cal F}^{++}_2(\dop,\bop,\zop)& =   \frak{D}_2(\dop,\bop)\,,\hskip 3.5cm {\cal G}_2^{++}(\dop,\bop,\zop,\iota) =0\,, \\ 
{\cal F}^{++}_3(\dop,\bop,\zop)& =   \frak{D}_3(\dop,\bop)- \beta \frak{D}_0(\dop,\bop)\,, \nn 
\end{align}
and after permuting $\bop \leftrightarrow\dop$ we also get the $(--)$ sector
\begin{align}
{\cal F}^{--}_1(\dop,\bop,\zop)& =  \frak{D}_1(\bop,\dop)+ \beta \frak{D}_0(\bop,\dop)\,,\hskip 1.5cm {\cal G}_1^{--}(\dop,\bop,\zop,\iota) =0\,, \nn \\ 
{\cal F}^{--}_2(\dop,\bop,\zop)& =   \frak{D}_2(\bop,\dop)\,,\hskip 3.5cm {\cal G}_2^{--}(\dop,\bop,\zop,\iota) =0\,, \\ 
{\cal F}^{--}_3(\dop,\bop,\zop)& =   \frak{D}_3(\bop,\dop)- \beta \frak{D}_0(\bop,\dop)\,, \nn 
\end{align}
where the entire differential functions are
\begin{align}
\frak{D}_0(\dop,\bop) := & 
\frac{1}{8 \bop \dop}
- \frac{4 + \dop}{16 \bop \dop^2} (e^\dop - 1) 
- \frac{2 + \dop }{16 \bop^2 \dop^2} (e^{-\bop} - 1) (e^\dop - 1) 
+\Big\{\bop \leftrightarrow -\dop\Big\} \,,\nn \\
\frak{D}_1(\dop,\bop) := &\frac{(\dop-\bop) (e^{-\bop} - 1) (e^\dop - 1) + \bop \dop (e^{-\bop} + e^\dop - 2)}{8 \bop^2 \dop^2} \,, \nn
\\ 
\frak{D}_2(\dop,\bop) := & -\frac{1 + b}{8 \bop^3 \dop^3} (e^{-\bop} - 1) (e^\dop - 1) 
- \frac{4 - 2 \dop + \bop \dop }{16 \bop^2 \dop^3} (e^\dop - 1) 
+ \frac{4 + \bop \dop}{32 \bop^2 \dop^2} +\Big\{\bop \leftrightarrow -\dop\Big\} \,, \nn
\\
\frak{D}_3(\dop,\bop) := & 
\frac{(1 + d) (3 b-1)}{16 b^2 d^2}
+ \frac{b -d-1}{8 (b - d)^2 d^2}
+ \frac{3 d^2 + 2 d + 2b + b d}{8 b^2 d^3} e^{-b}  \nn \\
& + \frac{b e^{d - b}}{8 (b - d)^2 d^3}
- \frac{(1 + b) (2 b + 3 d)}{8 b^3 d^2} e^{d - b} +\Big\{\bop \leftrightarrow -\dop\Big\}\,.
\end{align}
Note that these $R^2$ contributions in the same-helicity sector has only one purpose, namely to subtract out the unwanted contact-term contributions that the linear-in-$R$ operators generate. Thus one should not attribute too much significance to these operators, since the same-helicity amplitude has no true dependence on them.   

We now give the opposite-helicity differential functions, which we split up into seven entire functions for convenience
\begin{align} \label{Fpm_Operators}
{\cal F}^{-+}_1(\dop,\bop,\zop)& = \frak{A}_1(\dop,\bop,\zop) +  \frak{B}_1(\dop,\bop)+ \beta \frak{D}_0(\dop,-\bop) \,,\hskip 1.1cm {\cal G}_1^{-+}(\dop,\bop,\zop,\iota) =0\,, \nn \\ 
{\cal F}^{-+}_2(\dop,\bop,\zop)& = \frak{A}_2(\dop,\bop,\zop) +  \frak{B}_2(\dop,\bop) \,,\hskip 3.42cm {\cal G}_2^{-+}(\dop,\bop,\zop,\iota) =0\,,\\ 
{\cal F}^{-+}_3(\dop,\bop,\zop)& = \frak{A}_3(\dop,\bop,\zop) +  \frak{B}_3(\dop,\bop)- \beta \frak{D}_0(\dop,-\bop)\,, \nn 
\end{align}
where the main reason for the split is that the variable dependence and functional form are quite different. 
The seven entire differential functions are
\begin{align}
\frak{B}_3(\dop,\bop):= &-\frac{(\bop + \dop) (e^\bop - 1) (e^\dop - 1) - \bop \dop (e^\bop + e^\dop - 2)}{8 \bop^2 \dop^2} \,, \nn
\\ 
\frak{B}_2(\dop,\bop):= & \frac{2 - 2 \dop + \bop \dop }{16 \bop^3 \dop^3} (e^\bop - 1) (e^\dop - 1) 
- \frac{4 - 2 \dop + \bop \dop }{16 \bop^2 \dop^3} (e^\dop - 1) 
+ \frac{4 + \bop \dop}{32 \bop^2 \dop^2} +\Big\{\bop \leftrightarrow \dop\Big\} \,, \nn
\\
\frak{B}_1(\dop,\bop):= &  \frac{3 \dop - 4}{16 \bop \dop^2} 
+ \frac{3 \dop + 4 \bop - 2 \dop^2}{16 \bop^2 \dop^3}  (e^\bop - 1) (e^\dop - 1) 
+ \frac{(2 \bop + \dop) (1 - 2 \dop)}{8 \bop^2  \dop^3}(e^\dop - 1) \nn \\
&+ \frac{ 1 - e^{\dop - \bop}}{8 \bop^2 (\bop - \dop)^2}
+\Big\{\bop \leftrightarrow \dop\Big\} \,.
\end{align}
Note that $\frak{B}_3(d,p)=-\frak{D}_1(d,-b)$ and $\frak{B}_2(d,b)=(\frak{D}_2-\frak{D}_1/4-\frak{D}_0/2)(d,-b)$, which follows from the relations in \eqn{f++definitions}. Again, the purpose of these $\frak{B}_i$ functions is to subtract out the unwanted contact terms that are generated from the linear-in-Riemann interactions, and thus the resulting amplitude has no dependence on these. In turn, the more physically relevant contact terms in the amplitude originate from the $\frak{A}_i$ functions, which are
\begin{align}
\frak{A}_3(\dop,\bop,\zop):=&\bigg[ \frac{\zop^2 e^{\bop - \dop}}{4 \dop (\bop - \dop) (4 \dop^2 - \zop^2)^2}-\frac{e^{\bop + \dop}}{ 8\bop \dop \zop^2}  + \frac{e^{\bop + \dop + \zop} }{2 \zop (2 \bop + \zop) (2 \dop + \zop)^2} - \frac{(\zop - 2)e^{\bop + \dop + \zop}}{4 \zop^2 (2 \bop + \zop) (2 \dop + \zop)} \nn \\ & + \Big\{\bop \leftrightarrow \dop \Big\}\bigg] + \Big\{\zop \rightarrow -\zop \Big\}\,, \nn \\ 
\frak{A}_2(\dop,\bop,\zop):=& \bigg[\frac{(\bop-1) e^{\bop + \dop}}{8 \bop^2 \dop \zop^2}
+ \frac{e^{\bop - \dop}}{8 (\dop-\bop ) \dop^2 (4 \dop^2 - \zop^2)} 
- \frac{e^{\bop + \dop + \zop}}{2 \zop^3 (2 \bop + \zop) (2 \dop + \zop)} 
 + \Big\{\bop \leftrightarrow \dop \Big\}\bigg] \nn\\ &+ \Big\{\zop \rightarrow -\zop \Big\} \,, \nn
\\ 
\frak{A}_1(\dop,\bop,\zop) := & \bigg[
\frac{\dop e^{\bop- \dop}}{(\dop-\bop) (4 \dop^2 - \zop^2)^2}  
+ \frac{e^{\bop + \dop + \zop}}{4 \zop (2 \bop + \zop) (2 \dop + \zop)} 
- \frac{e^{\bop + \dop + \zop}}{2 \zop (2 \bop + \zop) (2 \dop + \zop)^2}
 + \Big\{\bop \leftrightarrow \dop \Big\}\bigg] \nn \\ & + \Big\{\zop \rightarrow -\zop \Big\} \,.
\end{align}
The symmetrization brackets at the end signify two rules that are implemented consecutively.

Importantly the $\frak{A}_i,\frak{B}_i$ and $\frak{D}_i$ are entire functions, thus they have no poles, and are well-defined differential operators, this can be made manifest after expanding out the exponential factors.  The first few orders of $\frak{D}_i$ are
\begin{align}
\frak{D}_1(\dop,\bop) &= \frac{1}{8} - \frac{5}{96} (b - d)  + \frac{1}{192} (3 b^2 - 4 b d + 3 d^2)+ \dots \,, \nn \\
\frak{D}_2(\dop,\bop) &= -\frac{1}{72} + \frac{5}{1152} (b - d) + \frac{1}{3840}(-4 b^2 + 5 b d - 4 d^2) + \dots \,,\nn \\
\frak{D}_3(\dop,\bop) &= -\frac{1}{8} + \frac{23}{480} (b - d) +\frac{1}{1440} (-19 b^2 + 28 b d - 19 d^2)+ \dots \,,\nn \\
\frak{D}_0(\dop,\bop) &=-\frac{1}{8} + \frac{3}{64} (b - d)  - \frac{5}{1152} (3 b^2 - 4 b d + 3 d^2)+ \dots
\end{align}
and we note that the leading terms cancel for the combinations $\frak{D}_1+\frak{D}_0$ and $\frak{D}_3-\frak{D}_0$, which corresponds to the absence of $S^4$ operators in the same-helicity sector for the $\beta=1$ prescription.

In the opposite-helicity sector, the first few orders of $\frak{A}_i$ are
\begin{align}
    \frak{A}_1(\dop,\bop,\zop) &= \frac{1}{12}+  \frac{\bop + \dop}{20} +  \frac{1}{360} (7 \bop^2 + 10 \bop \dop + 7 \dop^2) + \frac{\zop^2}{180} + \dots  \,,\nn \\
    \frak{A}_2(\dop,\bop,\zop) &= -\frac{1}{60}-  \frac{\bop + \dop}{90} -  \frac{1}{2520} (11 \bop^2 + 18 \bop \dop + 11 \dop^2) - \frac{\zop^2}{2520} + \dots \,, \\
    \frak{A}_3(\dop,\bop,\zop) &= -\zop^2 \left(\frac{1}{360}+  \frac{\bop + \dop}{504} + \frac{1}{10080}(8 \bop^2 + 14 \bop \dop + 8 \dop^2 + \zop^2) + \dots \right) \nn\,.
\end{align}
and likewise for $\frak{B}_i$
\begin{align}
    \frak{B}_1(\dop,\bop) &= \frac{1}{24} + \frac{1}{120} (b + d)  + \frac{1}{720} b d  -\frac{1}{40320}(b + d) (9 b^2 + 14 b d + 9 d^2)+ \dots  \,,\nn \\
    \frak{B}_2(\dop,\bop) &= \frac{5}{288} + \frac{7}{1152} (b + d)+ \frac{1}{11520} (18 b^2 + 25 b d + 18 d^2) + \dots \,, \\
    \frak{B}_3(\dop,\bop) &= -\frac{1}{8} - \frac{5}{96} (b + d)  - \frac{1}{192} (3 b^2 + 4 b d + 3 d^2) + \dots \nn\,.
\end{align}
We note a curious feature: all coefficients in the series expansion of $\{\frak{A}_1,-\frak{A}_2,-\frak{A}_3\}$ are positive.  Likewise, all series coefficients of $\{\frak{B}_2,-\frak{B}_3\}$ are positive; however, the first few coefficients of $\frak{B}_1$ are positive, then they become negative. If we instead consider $\frak{B}_1(d,b)+\frak{D}_0(d,-b)$, which is the relevant combination for $\beta=1$ prescription, then all its coefficients are negative. In the light of this observation, it would be interesting to see if the signs of the Wilson coefficients can be deduced from some kind of EFT positivity argument.   

On-shell the differential operator $\zop$ becomes $z=2\omega |a|$, and as hinted at previously the positive powers $z^n$ in the tree amplitude can be expected to be modified by loop corrections. Their Wilson coefficients should be more properly fixed by loop-level matching to the Teukolsky equation, which is beyond the scope of the current paper. However, we can ameliorate this by considering those Wilson coefficients that are not proportional to $z$, which we can conveniently pick out by setting $|a| \to 0$, or rather at the level of the spin Casimir $S^2\to 0$, which requires $S^\mu$ to be a complex null vector. 

Even so, an infinite number of $\zop$-independent operators are required to reproduce the all-spin Compton amplitude even without the $z$ contact terms.
Let us use the following short-hand notation for spatial derivatives acting the curvature tensors: 
\be
E^{\pm, n}_{\mu \nu} := \Big(\mp \frac{i}{m} S \cdot \nabla\Big)^n E^{\pm}_{\mu \nu}\,,
\ee 
then the first few terms in $\cL_{R^2}$, which survive the $S^2\to 0$ limit, are 
\begin{align}
    m^2 \cL_{R^2}\Big|^{\frak{A}_1}_{S^2\to 0} = &\frac{8}{3} E^{-}_{SS} E^{+}_{SS}+\frac{16}{5 } E^{-}_{SS} E^{+,1}_{SS} +\frac{8}{9} E^{-,1}_{SS}  E^{+,1}_{SS}    + \frac{56}{45} E^{-}_{SS} E^{+,2}_{SS}   + {\cal O}(S^7)+{\rm h.c.}\,, \nn \\
    m^2\cL_{R^2}\Big|^{\frak{B}_1}_{S^2\to 0}  = &\frac{4}{3} E^{-}_{SS} E^{+}_{SS}+\frac{8}{15} E^{-}_{SS} E^{+,1}_{SS} +\frac{2}{45} E^{-,1}_{SS}  E^{+,1}_{SS}       + {\cal O}(S^7)+{\rm h.c.}\,,  \\
    m^2\cL_{R^2}\Big|^{\frak{D}_0}_{S^2\to 0}  = &-4 E^{-}_{SS} E^{+}_{SS}-3 E^{-}_{SS} E^{+,1}_{SS}  -\frac{5}{9} E^{-,1}_{SS}  E^{+,1}_{SS}   - \frac{5}{6} E^{-}_{SS} E^{+,2}_{SS}\nn \\ &
    -4 E^{+}_{SS} E^{+}_{SS}-\frac{3}{32} E^{+}_{SS} E^{+,1}_{SS}  -\frac{5}{9} E^{+,1}_{SS}  E^{+,1}_{SS}   - \frac{5}{6} E^{+}_{SS} E^{+,2}_{SS}
    + {\cal O}(S^7)+{\rm h.c.}\,, \nn \\
    m^2\cL_{R^2}\Big|^{\frak{D}_1}_{S^2\to 0}  = & 4  E^{+}_{SS} E^{+}_{SS}+\frac{10}{3} E^{+}_{SS} E^{+,1}_{SS}  +E^{+,1}_{SS}  E^{+,1}_{SS}   + \frac{4}{3} E^{+}_{SS} E^{+,2}_{SS}
    + {\cal O}(S^7)+{\rm h.c.}\,,\nn
\end{align}
where we now multiplied out the $2^5$ factor that appeared in \eqn{ActionPolarS}. 

Summing over all above contributions allows us to write out the first few terms in the $R^2$ operator, for which we use the prescription $\beta=1$ and set the Casimir $S^2$ to zero, giving
\begin{align}
    m^2 \cL_{R^2}\Big|^{\beta=1}_{S^2\to 0} = &~  \frac{11}{15} E^{-}_{SS} E^{+,1}_{SS} + \frac{17}{45} E^{-,1}_{SS}  E^{+,1}_{SS} + \frac{37}{45} E^{-}_{SS} E^{+,2}_{SS}  \nn \\ &
    +\frac{1}{3} E^{+}_{SS} E^{+,1}_{SS}  +\frac{1}{9} E^{+,1}_{SS}  E^{+,1}_{SS}   + \frac{1}{6} E^{+}_{SS} E^{+,2}_{SS}
    + {\cal O}(S^7)+{\rm h.c.}\,.
\end{align}
For this prescription, all the $S^4R^2$ operators cancel out and the corrections start at the $S^5$ spin-multiple order~\cite{Ben-Shahar:2023djm,Scheopner:2023rzp}. As previously stated, we refrain from attributing physical meaning to the Wilson coefficients of these operators, since they are polluted by the $\frak{B}_i$ and $\frak{D}_i$ terms whose only purpose is to remove unwanted contributions coming from the linear-in-$R$ worldline action. The somewhat more meaningful operators are those coming from the $\frak{A}_i$ entire functions.

\subsection{Dissipative $R^2$ operators}
We now want to write up Riemann-squared interactions for dissipative effects, which seem to appear naturally both in the Teukolsky results of \rcite{Bautista:2022wjf} and in the higher-spin Compton amplitude of \rcite{Cangemi:2023bpe}. These are the terms proportional to $\eta$ in \eqn{Compton_from_HS}. However, there are a number of potential issues with these terms that warrant some caveats. 
 Most pressingly, it is unclear if these terms should be interpreted as tree-level contributions, or if they should be generated from loop effects in an EFT framework. In \rcite{Bautista:2022wjf} they were included in the tree amplitude, but in later work \cite{Bautista:2023sdf} it was argued that they are loop corrections. They start at order $S^5$ in the spin-multiple expansion, and as seen in \eqn{Compton_from_HS} the first term is proportional to $z (w^2 - z^2)^2$ and all higher-order terms are odd in $z$. However, from \rcite{Bautista:2022wjf} it is clear that all such terms originate from an overall square-root factor prior to taking the week-coupling limit,
\be \label{Square-root-issue}
 -2i \omega \sqrt{(Gm)^2-|a|^2} ~~\stackrel{G\to 0}{\longrightarrow}~~ 2 \omega |a| =z\,,
\ee
where we have suppressed the $(w^2 - z^2)^2$ factor, as we focus on the linear-in-$z$ term. Let us mention two features that support a loop interpretation. Firstly, the square root factor is suggestive of a loop effect, both since it contains $G$ to all orders, and since such a non-rational function of spin is unlikely to appear in a tree-level higher-spin QFT Lagrangian. Secondly, as written above\footnote{Note that factor \eqref{Square-root-issue} is presented on a suggestive form for the purpose of the argument. It can equally well be presented as $i\epsilon\sqrt{1-\chi^2}$, where $\chi=|a|/(Gm)$ and  $\epsilon=2Gm\omega$, see \rcite{Bautista:2022wjf}. From a worldline-EFT perspective the square root is simply a number that does not necessarily suggest a loop interpretation; although, the overall $\epsilon$ now stands out instead.}, the variables are not yet in strictly classical combinations; this can be fixed by moving the frequency $\omega$ inside the square root, giving $ -i\sqrt{\epsilon^2-z^2}$, where $\epsilon=2Gm\omega$. If this is a reasonable rewriting, then the contribution depends on $|\omega|$ rather than $\omega$, which is a non-local kinematic variable (time derivative) that can only be generated by loop effects in an EFT. Having stated these warnings, we will in the remaining part of this section entertain the possibility that the linear-in-$z$ terms of \eqn{Compton_from_HS} justifies the inclusion of dissipative operators in the tree-level action.

A final warning, while the left-hand side of \eqn{Square-root-issue} is dissipative because of the imaginary unit, the right-hand side has lost this feature because of the extra $\sqrt{-1}$ that appears. This happens because $G\to 0$ limit is outside the physical region, since it corresponds to a super-extremal Kerr black hole \cite{Siemonsen:2019dsu,Bautista:2022wjf}. Thus, the dissipative properties (or reality properties) are messed up by this limit. Nevertheless, we will in this section try to make sense, from a Lagrangian perspective, of these linear-in-$z$ terms, which we will continue to call ``dissipative'' despite this being a slight misnomer.

In general, dissipative or inelastic effects in a worldine action break time reversal symmetry (and thus break $CPT$) by violating the Bose symmetry of the gravitons for terms odd in $\omega$. This makes it difficult to formulate such interaction operators using standard worldline quantum fields. One option for dealing with this is to introduce the dissipative effects following \rcites{Goldberger:2005cd, Porto:2007qi, Goldberger:2019sya,
Goldberger:2020wbx, Goldberger:2020fot, Jones:2023ugm}, by implicitly adding microscopic degrees of freedom $Q_{\mu \nu}^\pm$ on the worldline which couple to the macroscopic degrees of freedom, e.g. $E_{\mu \nu}^\pm$, with additional differential operators acting on the fields.
However, here we prefer to instead  use the Schwinger-Keldysh approach of doubling a field $\phi \to \{\phi^{\rm in}, \phi^{\rm out}\}$ by keeping track of the in versus out states. Specifically, we apply this to the curvature fields 
\be
E^{\pm}_{\mu \nu} \to \big\{E^{\pm\,{\rm in}}_{\mu \nu}, E^{\pm\,{\rm out}}_{\mu \nu}\big\}  \,. 
\ee
For the Compton process, this is is equivalent to above mentioned approach, since we may think of $E^{\pm{\rm in}}_{\mu \nu} \sim Q_{\mu \nu}^\pm$, and then $E^{\pm\,{\rm out}}_{\mu \nu}$ is the standard curvature tensor.

Similarly to the conservative sector of the worldline, we can split the dissipative $R^2$ operators into two natural classes: those that vanish for polar scattering and those that contribute. Thus, the dissipative operators for polar-vanishing and polar-contributing cases are
\begin{align} \label{ActionPolarS_dissipative}
{\cal L}_{R^2}^\text{dis-pv}&=\frac{2^5}{m^2}\big(E^{-\rm out}_{S\mu}  {\cal O}_{-+}^{\mu \nu} E^{+\rm in}_{S\nu}
+E^{+\rm out}_{S\mu}  {\cal O}_{+-}^{\mu \nu} E^{-\rm in}_{S\nu}
+ E^{+\rm out}_{S\mu} {\cal O}_{++}^{\mu \nu} E^{+\rm in}_{S\nu}
+ E^{-\rm out}_{S\mu}  {\cal O}_{--}^{\mu \nu} E^{-\rm in}_{S\nu}\big)\,, \nn \\
{\cal L}_{R^2}^\text{dis-pc} &= \frac{2^5}{m^2}g^{\rho \sigma} \Big(E^{-\rm out}_{\mu \rho}  \widetilde {\cal O}_{-+}^{\mu \nu}  E^{+\rm in}_{\nu \sigma}
+E^{+\rm out}_{\mu \rho}  \widetilde {\cal O}_{+-}^{\mu \nu}  E^{-\rm in}_{\nu \sigma}
+E^{+\rm out}_{\mu \rho} \widetilde {\cal O}_{++}^{\mu \nu}  E^{+\rm in}_{\nu \sigma}
+ E^{-\rm out}_{\mu \rho} \widetilde {\cal O}_{--}^{\mu \nu}  E^{-\rm in}_{\nu \sigma}\Big)\,,
\end{align}
Sandwiched between the curvature tensors we use the following differential operators
\begin{align}\label{Ooperators_dissipative}
 {\cal O}_{\pm\pm}^{\mu \nu} &:=  \frac{i \eta |S|}{m}  \Big[\Big(S^\mu S^\nu
- \frac{1}{2} g^{\mu\nu} S^2\Big) {\cal H}^{\pm \pm}_1 - i \frac{ {\cal H}^{\pm \pm}_2}{m}\Big(S^\nu \nabla^\mu- g^{\mu\nu}  S\cdot \nabla\Big) -\frac{1}{2}g^{\mu\nu} S^2 {\cal H}^{\pm \pm}_3 \Big] \hat p \cdot \nabla\,,  \nn \\
\widetilde {\cal O}_{\pm \pm}^{\mu \nu}& :=    \frac{i \eta |S|}{m}  \Big[\frac{1}{2}g^{\mu\nu} S^4 {\cal K}_1^{\pm \pm}  + i S^2 \frac{{\cal K}_2^{\pm \pm}}{m} (S^\nu \nabla^\mu- g^{\mu\nu}  S\cdot \nabla)\Big] \hat p \cdot \nabla \,,
\end{align}
which have essentially the same structure as the conservative $O_{\pm\pm}^{\mu \nu}$ operators in \eqns{Ooperators}{OtidleOperator}, except that we multiplied by an overall power of $\eta \vartheta$ operator, where $\eta=\pm1$ keeps track of the boundary conditions of the dissipative effects (absorption vs. emission at the black hole horizon).

The ${\cal H}_k^{\pm\pm}$ and ${\cal K}_k^{\pm\pm}$ are entire functions that, similar to the conservative sector, depend on three and four differential operators, respectively, 
\be
{\cal H}_k^{\pm\pm}={\cal H}_k^{\pm\pm}(d,b,\vartheta)\,,~~~~~~~  {\cal K}_k^{\pm\pm}={\cal K}_k^{\pm\pm}(d,b,\vartheta,\iota)\,,~~~~~~~  
\ee
where $d,b,\vartheta$ were defined \eqn{differentialOps} and the fourth differential operator $\iota$ is defined in \eqn{iotaDef}. We impose that the hermitian conjugate of the ${\cal H}$ and ${\cal K}$ functions reverses the superscript signs, specifically
\be
{\cal H}^{+-}_k = ({\cal H}^{-+}_k)^\dagger\,,~~ {\cal H}^{--}_k = ({\cal H}^{++}_k)^\dagger\,,~~ {\cal K}^{+-}_k = ({\cal K}^{-+}_k)^\dagger\,,~~ {\cal K}^{--}_k = ({\cal K}^{++}_k)^\dagger\,.
\ee

We can now evaluate the Compton matrix elements of the operators. For the polar-vanishing operators they are
\begin{align}
\frac{2^5}{m^2}\Big\langle  E^{-\rm out}_{S\mu} {\cal O}_{-+}^{\mu \nu} E^{+\rm in}_{S\nu} \Big\rangle &=\frac{(v \cdot \chi)^4}{2(v \cdot q_\perp)^4}\eta z (w^2-z^2)( w^2 {\cal H}^{-+}_1+ w {\cal H}^{-+}_2+z^2 {\cal H}^{-+}_3)\,, \nn \\
\frac{2^5}{m^2}\Big\langle    E^{+\rm out}_{S\mu} {\cal O}_{++}^{\mu \nu} E^{+\rm in}_{S\nu} \Big\rangle & =\frac{(v \cdot X)^4}{2(v \cdot q_\perp)^4} \eta z (u^2-z^2)(u^2 {\cal H}^{++}_1+ u {\cal H}^{++}_2+z^2 {\cal H}^{++}_3)\,, \nn \\
\frac{2^5}{m^2}\Big\langle  E^{-\rm out}_{S\mu} {\cal O}_{--}^{\mu \nu} E^{-\rm in}_{S\nu} \Big\rangle &= \frac{(v \cdot \bar X)^4}{2(v \cdot q_\perp)^4} \eta z (\bar u^2-z^2)( \bar u^2 {\cal H}^{--}_1+\bar u  {\cal H}^{--}_2+ z^2 {\cal H}^{--}_3)\,,
\end{align}
and for the polar-contributing operators we have
\begin{align}
\frac{2^5}{m^2}\Big\langle  E^{-\rm out}_{S\mu} {\cal O}_{-+}^{\mu \nu} E^{+\rm in}_{S\nu} \Big\rangle &= \frac{(v \cdot \chi)^4}{2(v \cdot q_\perp)^4} \eta z ( z^4 {\cal K}^{-+}_1+ w z^2 {\cal K}^{-+}_2)\,, \nn \\
\frac{2^5}{m^2}\Big\langle  E^{+\rm out}_{S\mu} {\cal O}_{++}^{\mu \nu} E^{+\rm in}_{S\nu} \Big\rangle & = \frac{(v \cdot X)^4}{2(v \cdot q_\perp)^4}\eta z  (z^4 {\cal K}^{++}_1+ u z^2 {\cal K}^{++}_2)\,, \nn \\
\frac{2^5}{m^2}\Big\langle   E^{-\rm out}_{S\mu} {\cal O}_{--}^{\mu \nu} E^{-\rm in}_{S\nu} \Big\rangle &= \frac{(v \cdot \bar X)^4}{2(v \cdot q_\perp)^4} \eta z ( z^4 {\cal K}^{--}_1+ \bar u z^2  {\cal K}^{--}_2)\,.
\end{align}
Above, the entire functions now depend on the kinematic variables as ${\cal H}^{\pm\pm}_k={\cal H}^{\pm\pm}_k(\frac{x+y}{2},\frac{x-y}{2},z)$ 
and 
${\cal K}^{\pm\pm}_k={\cal K}^{\pm\pm}_k(\frac{x+y}{2},\frac{x-y}{2},z,q^2|a|^2)$. 
The ${\cal H}^{\pm\pm}_k$s and  ${\cal K}^{\pm\pm}_k$s can be made to satisfy the same hermiticity properties and $T$-symmetry as the ${\cal G}^{\pm\pm}_k$s; see \eqn{F_Psymmetry}, \eqref{F_Tsymmetry}, as well as \eqn{xyzSymetries}. 

After adding together ${\cal L}_{R^2}^\text{dis-pv}$ and ${\cal L}_{R^2}^\text{dis-pc}$, the total number of independent dissipative $R^2$ operators  is given by the generating series
\be \label{OperatorCount_dissipative}
2\Big(\frac{t^4 + 2t^5}{(1 - t)^3 (1 + t)^2}+ \frac{t^4 + t^5}{(1 - t)^4 (1 + t)^3}\Big) = 2\big(2 t^4 + 5 t^5 + 10 t^6 + 17 t^7 + 26 t^8 + 38 t^9 + 52 t^{10}+\ldots \big)
\ee
where the two terms with different denominator factors again correspond to the split into polar-vanishing operators and  polar-contributing operators. The overall factor of 2 is again because the same-helicity and opposite-helicity sectors contribute equally to the count. Note that the Wilson coefficients are in principle allowed to be functions of the dimensionless quantity $|a|/(Gm)$, and thus we do not consider powers of this quantity as giving rise to independent operators. The count in \eqn{OperatorCount_dissipative} agrees with \rcite{Haddad:2023ylx}.

\vskip0.3cm
\noindent
{\bf Final results for dissipative $R^2$ operators:}\\
For the dissipative $R^2$ operators we need only one entire function to match the higher-spin QFT results \eqref{Compton_from_HS}, namely the function $\partial_z \tilde E$, which translated into the above basis of operators become
\be
{\cal H}_1^{-+} =-{\cal H}_3^{-+}= \left(\frac{2 e^{b - d} (d + \vartheta) + e^{b + d + \vartheta} (2 d (\vartheta - 1) + (\vartheta - 2) \vartheta)}{4 (b - d) \vartheta^3 (2 d + \vartheta)^2}+\big\{b \leftrightarrow d \big\} \right) + \big\{\vartheta \to -\vartheta \big\}\,,
\ee
and the remaining functions are taken to vanish
\begin{align}
&{\cal H}_2^{\pm \pm}=0\,, \nn \nn \\
&{\cal H}_k^{++}={\cal H}_k^{--}=0\,,\hskip1cm (k=1,2,3)
 \nn \\
&{\cal K}_k^{\pm\pm}=0\,.\hskip2.35cm (k=1,2)
\end{align}
Let us also plug in the definitions, and display the final result, the dissipative $R^2$ operators to all orders in spin, that matches \eqn{Compton_from_HS}, then becomes the simple result
\begin{align}
{\cal L}_{R^2}^\text{dis}&=2^5 \frac{i\eta }{m^3} |S| E^{-\rm out}_{SS}  {\cal H}^{-+}_1 \hat p\cdot \nabla  E^{+\rm in}_{SS}\,,
\end{align}
where we recall that ${\cal H}^{-+}_1$ is a differential operators acting on $E^\pm_{SS}$.

\section{Conclusion}
\label{sec_conclusion}

In this paper, we considered the spinning worldline action of~\rcites{Levi:2015msa,Ben-Shahar:2023djm}, which we combined with worldline-QFT methods~\cite{Mogull:2020sak,Haddad:2024ebn} to compute tree-level Compton amplitudes to all orders in the spin multipole expansion, re-summed into entire functions for both the same- and opposite-helicity sectors. We then compare the results to the all-orders-in-spin opposite-helicity Compton amplitude obtained in \rcite{Cangemi:2023bpe}, which was obtained from a higher-spin QFT framework. We work out the difference in terms of quadratic-in-Riemann operators ($R^2$) that are added to the worldline action. 

The spinning worldline action~\cite{Porto:2005ac,Porto:2006bt,
Porto:2008tb,Porto:2008jj,Levi:2008nh,Porto:2010tr,Porto:2010zg,
Levi:2010zu,Levi:2011eq,
Porto:2012as,Levi:2014gsa,
Levi:2014sba,Levi:2015msa,Levi:2015uxa,Levi:2015ixa, 
Levi:2016ofk,Maia:2017yok, 
Maia:2017gxn,
Levi:2020kvb,Levi:2020uwu,Liu:2021zxr} has been an important tool as an effective theory for modeling the dynamical behavior of rotating Kerr black holes. Describing the degrees of freedom of a Kerr black hole is considerably more complicated than the Schwarzschild case, and requires proper treatment of spin gauge symmetry~\cite{Levi:2015msa} (see also~\rcite{Ben-Shahar:2023djm}) such that the equations of motion preserve the physical degrees of freedom for a particle of constant spin magnitude and mass. Additionally, an infinite family of non-minimal interactions~\cite{Levi:2015msa} that are linear in the Riemann tensor are needed for describing Kerr at leading post-Minkowskian (PM) order~\cite{Vines:2017hyw}. Corrections that are non-linear in the Riemann tensor are on general grounds expected; however, it is known that up to $S^4$ the linear-in-$R$ terms seem sufficient for matching to GR predictions~\cite{Siemonsen:2019dsu,Ben-Shahar:2023djm, Scheopner:2023rzp}. 

The Compton process directly probes the quadratic-in-$R$ operators of the worldline, and serves as important input for binary black hole scattering and related computations of observables~\cite{Chen:2021kxt,Bautista:2023szu,Bautista:2024agp,Bohnenblust:2024hkw}. Both the Compton amplitude and binary 2PM scattering were computed from the worldline up to $S^4$ in~\rcite{Ben-Shahar:2023djm} confirming that $R^2$ operators are not needed to describe Kerr. Similar analysis was carried out in \rcite{Scheopner:2023rzp} for the Compton process, and in \rcite{Bautista:2024agp} for binary observables,  which both suggest that at both $S^5$ and $S^6$ orders the $R^2$ worldline corrections are needed for properly matching to Kerr black hole perturbation theory results obtained via the Teukolsky equation~\cite{Bautista:2022wjf,Bautista:2023sdf}. Note that the $R^2$ operators introduced in this paper are exclusively used to describe far-zone physics (see e.g.~\rcite{Bautista:2024agp}), thus they are not necessarily related to tidal Love numbers or near-zone physics. Of course, a complete EFT should also model near-zone effects \cite{Ivanov:2022qqt,Ivanov:2024sds,Saketh:2023bul,Correia:2024jgr}, which would require proper matching of loop corrections in our worldline description, which is beyond the scope of the current paper.

Returning to the explicit worldline calculations of this work, we have shown that one can compute the tree-level Compton process from the linear-in-$R$ worldline~\cite{Levi:2015msa} to all orders in spin, thus resuming both the opposite- and same-helicity amplitudes into entire functions. The entire functions are built out of exponential functions with two variables: powers of $e^{x/2}$ and $e^{y/2}$ multiplied by rational prefactors. The rational factors naively seem to contain poles, however these are always compensated by corresponding zeros in the numerators, thus guaranteeing that the functions are entire. The worldline entire functions were then contrasted with the entire functions obtained from a higher-spin QFT framework in \rcite{Cangemi:2023bpe}, suggested to model Kerr. We observed that the analytic structure differs significantly as the latter are built out of exponentials with three arguments that are twice as large: $e^x$, $e^y$ and $e^z$. Furthermore, beyond $S^4$ the mismatch becomes considerable, with the higher-spin results being significantly cleaner. Specifically, in the same-helicity sector it is well-known that both the higher-spin results~\cite{Johansson:2019dnu,Chen:2021kxt,Aoude:2022trd} and the Teukolsky results~\cite{Bautista:2022wjf} for Kerr background agree on the presence of a single exponential $e^y$ in the Compton amplitude, whereas the worldline beyond $S^4$ produces an increasingly large number of terms that have no apparent physical meaning for a Kerr black hole.
To construct the action that reproduces the desired amplitude we classified $R^2$ operators and identified maps, such as \eqn{O_to_F_ops}, between these operators and on-shell variables (see also appendix \ref{appendix_action_to_onshell}). The resulting counting of independent operators, which we derive from simple generating functions, agrees with~\rcite{Haddad:2023ylx}. We then wrote down the full family of $R^2$ operators needed for aligning the worldline action with the higher-spin Compton amplitude, by examining the difference between the naive amplitude generated by linear in $R$ terms and the amplitude~\cite{Cangemi:2023bpe} we want to reproduce. The resulting $R^2$ action is written to all orders in spin through compact notation that uses operator-valued entire functions.

The opposite-helicity higher-spin Compton amplitude of~\rcite{Cangemi:2023bpe} has many desirable properties making it a plausible candidate for describing tree-level Kerr interactions; most significantly, its classical limit reproduces (without finetuning) large portions of the far-zone Teukolsky solution computed in~\rcite{Bautista:2022wjf,privateBGKV,Bautista:2023sdf}. Specifically, it matches all rational contributions that are not proportional to polygamma functions (as defined in~\rcite{Bautista:2022wjf}, where such non-rational terms were tagged by an auxiliary parameter $\alpha$). While there is a well-known mathematical ambiguity in uniquely splitting rational versus non-rational contributions, the non-trivial automatic matching between the higher-spin Compton amplitude and Teukolsky results for $\alpha=0$ is very suggestive. We only consider the case $\alpha=0$ in this paper, but we note that the dependence on $\alpha$ seem to always come proportional to powers of $z=2\omega |a|$, meaning that they vanish if the spin Casimir $S^2$ is set to zero. On general grounds, we expect that such $z$-dependent terms can be highly sensitive to loop corrections, meaning that their Wilson coefficients can only be fixed with confidence after considering loop-level matching to Teukolsky equation, including UV counter terms and renormalization in the worldline or EFT action. However, we note that our results should be more robust in the $z\to 0$ limit\footnote{By dimensional analysis it follows that a $L$-loop contribution can only mix with a tree-level contribution if there exists a Wilson coefficient proportional to $|S|^L/G^L$. Thus only Casimirs are affected by such mixing ambiguities.} (keeping spin-vector quantities $x,y,w,u$ fixed), which corresponds to analytic continuation of the spin vector into the complex plane such that its magnitude becomes null (i.e. Casimir $S^2=0$). We leave it to upcoming work to check the robustness of our results, by computing the Compton amplitude at loop level using worldline and EFT methods, and matching the Wilson coefficients to the predictions of the Teukolsky equation.

\section*{Acknowledgements}

We thank Dogan Akpinar, Fabian Bautista, Lara Bohnenblust, Marco Chiodaroli, Gang Chen, Paolo Di Vecchia, Alfredo Guevara, Alex Ochirov, Julio Parra-Martinez, Paolo Pichini, Radu Roiban, Trevor Scheopner, Oliver Schlotterer, Chia-Hsien Shen, Evgeny Skvortsov, Fei Teng for useful discussions. The research is supported by the Knut and Alice Wallenberg Foundation under grants KAW 2018.0116 and KAW 2018.0162. The research of M. B. S. is supported by the Knut and Alice Wallenberg Foundation under grant KAW 2023.0490.
H.J. acknowledges hospitality and support from the Munich Institute for Astro-, Particle and BioPhysics (MIAPbP), which is funded by the Deutsche Forschungsgemeinschaft (DFG, German Research Foundation) under Germany's Excellence Strategy – EXC-2094 – 390783311.
This research was supported in part by grant NSF PHY-2309135 to the Kavli Institute for Theoretical Physics (KITP). H.J. thanks KITP for the hospitality during the completion of this work.

\appendix
\section{Conventions and transformation properties}
\label{appendix_A}
In our calculations, we use incoming plane wave factor with phase
\begin{equation}
    e^{ik\cdot x}
\end{equation}
and the metric is mostly minus,
\begin{equation}
    \eta_{\mu \nu} = \textrm{diag}(+1,-1,-1,-1)\,.
\end{equation}
The Levi-Civita tensor $\epsilon_{\mu\nu\rho\sigma}$ is totally antisymmetric, and includes the appropriate power of the metric determinant, normalized as
\begin{equation}
    \epsilon_{0123} = \sqrt{-g}= 1+ {\cal O}(\kappa)\,.
\end{equation}

The worldline fields and operators transform non-trivially under the discrete $P$ and $T$ symmetries, which we summarize in the following table:
\begin{center}
\begin{tabular}{ |c||c|c| }
       \hline
       &  $T$-even & $T$-odd \\ 
    \hline
       \hline
$P$-even & $\dot x_0,p_0,\zop,E_{ij},\Lambda_{iI}$    & $x_0,\tau,\Omega_{ij},S_{ij},S_i$ \\ 
       \hline
$P$-odd & $x_i,\dop,\bop$                & $\dot x_i,p_i,B_{ij}$ \\ 
       \hline
\end{tabular}
\end{center}
where $i,j$ are the spatial indices and $0$ is time.  The $\dop,\bop,\zop$ symbols are differential operators, which we recall are defined as
\be
\dop := -\frac{i}{m} S \cdot \nabla \,,~~~~~~\bop :=  \overset{\leftarrow}{\nabla} \cdot S  \frac{i}{m} \,,~~~~~~   \zop := -\frac{2i}{m} |S|\, \hat{p}\cdot \nabla\,.~~
\ee
It is also convenient to state the $P$ and $T$ transformations rules for tensors such that Lorentz covariance is manifest:
\begin{alignat}{2} \label{eq:tensortrfs}
&x^{\mu}\stackrel{P}{\longrightarrow} x_{\mu}\,,~~~ &&x^{\mu}\stackrel{T}{\longrightarrow} -x_{\mu}\,,\nn \\
&p_{\mu}\stackrel{P}{\longrightarrow} p^{\mu}\,,~~~ &&p_{\mu}\stackrel{T}{\longrightarrow} p^{\mu}\,,~~(\text{same for}~\dot x_\mu, i\nabla_\mu)\nn \\
&S_{\mu}\stackrel{P}{\longrightarrow} -S^{\mu}\,,~~~ &&S_{\mu}\stackrel{T}{\longrightarrow} S^{\mu}\,,\nn \\
&S_{\mu \nu}\stackrel{P}{\longrightarrow} S^{\mu \nu}\,,~~~ &&S_{\mu \nu}\stackrel{T}{\longrightarrow} -S^{\mu \nu} \,,\nn \\
&\Omega^{\mu \nu}\stackrel{P}{\longrightarrow} \Omega_{\mu \nu}\,,~~~ &&\Omega^{\mu \nu}\stackrel{T}{\longrightarrow} -\Omega_{\mu \nu} \,,\nn \\
&\Lambda_{\mu I}\stackrel{P}{\longrightarrow} \Lambda^{\mu I}\,,~~~ &&\Lambda_{\mu I}\stackrel{T}{\longrightarrow} \Lambda^{\mu I} \,, \nn \\
&E_{\mu \nu}\stackrel{P}{\longrightarrow} E^{\mu \nu}\,,~~~ &&E_{\mu \nu}\stackrel{T}{\longrightarrow} E^{\mu \nu} \,, \nn \\
&B_{\mu \nu}\stackrel{P}{\longrightarrow} -B^{\mu \nu}\,,~~~ &&B_{\mu \nu}\stackrel{T}{\longrightarrow} -B^{\mu \nu} \,,\nn \\
&E^{\pm}_{\mu \nu}\stackrel{P}{\longrightarrow} E^{\mp \mu \nu}\,,~~~ &&E^{\pm}_{\mu \nu}\stackrel{T}{\longrightarrow} E^{\pm \mu \nu}\,,
\end{alignat}
Note that the $T$ transformation acts by complex conjugation on all $i$'s in the action, according to standard conventions. Hence, the helicity of the curvature tensors $E^{\pm}$ is not flipped. 

In \Tab{tab:Bose&CC_symmetry}, we give some useful transformation properties of the variables in the Compton amplitude under the discrete relabeling symmetries (Bose symmetry) and complex conjugation. 
\begin{center}
\begin{table}[h]
    \centering
\begin{tabular}{ ||c||c|c|c|| }
   \hline
    & $(1\leftrightarrow 2)$  & $ (3\leftrightarrow 4)$ & c.c. \\
   \hline   \hline
     $ x=a{\cdot}q_\perp$ & $x\longrightarrow x$ & $x \longrightarrow -x$ &  $x \longrightarrow -x$ \\
       \hline
     $ y=a{\cdot}q$ & $y\longrightarrow y$ &  $y \longrightarrow y$ & $y \longrightarrow -y$ \\  
    \hline
     $ z=|a| v{\cdot}q_\perp$ & $z\longrightarrow -z$ &  $z \longrightarrow-z$  &  $z \longrightarrow z$  \\
       \hline
     $w=\frac{\langle 3|a|4]}{\langle 3|v|4]} v\cdot q_\perp$ & $w\longrightarrow w$ &  $w \longrightarrow -\bar w$ &  $w \longrightarrow -\bar w$ \\
       \hline
     $u=\frac{[3|va|4]}{[34]} v\cdot q_\perp$ & $u\longrightarrow u$  &  $u \longrightarrow   u$ &  $u \longrightarrow- \bar  u$ \\
       \hline
     coeff. $c$ &$c\longrightarrow c$ & $c \longrightarrow c$ &  $c \longrightarrow c^*$ \\
       \hline
\end{tabular}
    \caption{Transformation properties of variables under permutation symmetry, or complex conjugation of the amplitude.  The amplitude (as a whole, for all sectors) must be invariant under permutation symmetry and complex conjugation; that is, these operations will always transform an amplitude into an amplitude.  We see that invariance under $(1\leftrightarrow 2)$ symmetry requires $A(1,2,3,4)$ to be $z$-even. Thus from a formal perspective we must break the Bose symmetry $(1\leftrightarrow 2)$ to allow for $z$-odd contributions.  }
    \label{tab:Bose&CC_symmetry}
\end{table}
\end{center}

The discrete $C,P,T$ transformations have actions on the Compton scattering amplitudes with states $(1,\bar 2,3^\pm,4^\pm)$, where we put a bar on the black hole 2 to, in principle, allow for it to be charged. The transformations are then given by
\begin{align}
C M(1,\bar 2,3^\pm,4^\pm)&= M(\bar 1,2,3^\pm,4^\pm)\,,\nn \\
P M(1,\bar 2,3^\pm,4^\pm)&= M(1, 2,3^\mp,4^\mp)\Big|_{a\to-a}\,,\nn \\
T M(1,\bar 2,3^\pm,4^\pm)&= \Big[M(1,\bar 2,3^\mp,4^\mp)\Big]^* \Big|_{v\to-v,k\to-k}\,, \nn \\
CPT M(1,\bar 2,3^\pm,4^\pm)&= \Big[M(\bar 1, 2,3^\pm,4^\pm)\Big]^*\,.
\end{align}

\section{Relation between curvature operators and on-shell variables}\label{appendix_action_to_onshell}
We here demonstrate some useful identities between the linearized on-shell fields of the worldline and the on-shell variables $\{x,y,z, w, u, q^2, v\cdot q_\perp, v\cdot \chi,  v\cdot X\}$ used for the opposite- and same-helicity Compton amplitudes in previous sections.

Working with linearized fields, we can write the electric and magnetic parts of the Riemann tensor as
\begin{align}
E_{\mu \nu} &:= v^\rho v^\sigma R_{\mu \rho \nu \sigma} \to -\frac{1}{2} E_\mu E_\nu \,,\nn \\
B_{\mu \nu} &:= \frac{1}{2}v^\rho v^\sigma {\epsilon_{\nu\sigma}}^{ \kappa \lambda}  R_{\mu \rho \kappa \lambda} \to  -\frac{1}{2} E_{\mu} B_{\nu}\,,
\end{align}
where we introduced electric and magnetic vector fields related to an abelian field strength
\begin{align}
E_\mu &:= v^\nu F_{\mu\nu} = v^\nu(\partial_\mu A_\nu - \partial_\nu A_\mu) \,, \nn \\
B^\mu & := \frac{1}{2}v_\nu \epsilon^{\mu \nu \rho \sigma}  F_{\rho \sigma} =   \frac{1}{2}\epsilon^{\mu \nu \rho \sigma} v_\nu (\partial_\rho A_\sigma - \partial_\sigma A_\rho) \,,
\end{align}
and they satisfy the quadratic relation $B^\mu B^\nu=-E^\mu E^\nu$. Thus, we are making use of the double copy~\cite{Bern:2010ue} for linearized fields. 

It is more convenient to work with the (anti)-self-dual fields 
\begin{align}
E^{\pm}_\mu&:= \frac{1}{2}(E_\mu \pm i B_\mu)\,, \nn \\
E^{\pm}_{\mu \nu}&:= \frac{1}{2}(E_{\mu \nu} \pm i B_{\mu \nu}) \to- \frac{1}{2} E^{\pm}_\mu E^{\pm}_\nu \,,
\end{align}
which are closely related to the helicity states. For on-shell states $A_\mu(k)=\varepsilon_\mu^{\pm}(k)$, $h_{\mu\nu}(k)=\varepsilon_\mu^{\pm}(k)\varepsilon_\nu^{\pm}(k)$, these vector fields evaluate to
\begin{align}
E^+_\mu(k) &= -\frac{i}{2\sqrt{2}} [k|v\sigma_\mu |k] \,, \nn \\
E^-_\mu(k) &= -\frac{i}{2\sqrt{2}} \langle k| \sigma_\mu v |k \rangle \,,
\end{align}
assuming the helicity matches the (anti)-self-duality, otherwise one gets zero.

We can now study the Lorentz invariant elementary building blocks that are relevant for respective helicity sectors. In the (++) sector, we can evaluate the linearized curvature fields by saturating the free indices with $a^\mu$ or $q^\mu=(k_3+k_4)^\mu$. We focus on the spin-1 fields $E^\pm_\mu$, since  the building blocks for spin-2 fields then follow from the double copy. 
We have three independent elementary contractions, which can be chosen as
\begin{align} \label{eq:3bulidingblocks}
(E_3^+ \cdot a) (E_4^+ \cdot a)&= -[34]^2\frac{u^2- z^2}{8 (v\cdot q_\perp)^2}\,,
\nn \\ 
(E_3^+ \cdot a) (E_4^+ \cdot q)+(E_3^+ \cdot q) (E_4^+ \cdot a) &= \frac{1}{4}[34]^2(u-y)\,, \\ 
E_3^+ \cdot E_4^+ &= -\frac{1}{4}[34]^2\,.\nn 
\end{align}
Naively, we would have expected more, since one can also have
\begin{align}
(E_3^+ \cdot q) (E_4^+ \cdot q) &= -[34]^2\frac{(v\cdot q_\perp)^2+ q^2}{8}\,,\nn \\ 
(E_3^+ \cdot a) (E_4^+ \cdot q)-(E_3^+ \cdot q) (E_4^+ \cdot a)& = \frac{1}{4}[34]^2 x\,, 
\end{align}
however, these two are not independent since one can attribute the factors that multiply $[34]^2$ as coming from the differential operators in \eqn{differentialOps} that acts on the curvature fields. Thus we ignore these two to avoid over-counting terms in the worldline action. 

In the $(-+)$ sector the details are almost identical to the above discussion; however, it is convenient to use $q_\perp^\mu=(k_4-k_3)^\mu$ instead of $q^\mu$. Again, we have the three independent elements
\begin{align}
(E_3^- \cdot a) (E_4^+ \cdot a)&=\frac{(v \cdot \chi)^2}{8 (v \cdot q_{\perp})^2} (w^2-z^2)\,,
\nn \\ 
(E_3^- \cdot a) (E_4^+ \cdot q_{\perp})+(E_3^- \cdot q_{\perp}) (E_4^+ \cdot a) &=- \frac{1}{4}(v \cdot \chi)^2 (w -x)\,,
\nn \\ 
E_3^- \cdot E_4^+ &=\frac{1}{4} (v \cdot \chi)^2\,,
\end{align}
and two candidates that are not independent
\begin{align}
(E_3^- \cdot q_{\perp}) (E_4^+ \cdot q_{\perp}) &=-\frac{1}{4}\frac{(v \cdot \chi)^2 q^2}{2}\,,\nn \\ 
(E_3^- \cdot a) (E_4^+ \cdot  q_{\perp}) -(E_3^- \cdot q_{\perp}) (E_4^+ \cdot a) &=- \frac{1}{4}(v \cdot \chi)^2 y \,.
\end{align}
Thus, for spin-1 fields, we conclude that there are exactly $3+3$ independent quadratic-in-curvature elementary building blocks in the same- and opposite-helicity sectors, respectively. [The non-elementary ones can be obtained by applying the differential operators in \eqn{differentialOps}.] It is no surprise that this exactly matches the polynomial basis of the $w,u$ variables; namely, we allow these variables to appear at most to second power, $\{1,w,w^2\}$ and $\{1,u,u^2\}$, to avoid spurious poles. 

To get the more relevant spin-2 result, we can invoke the double copy, which amounts multiplying out two copies of the above results, which in the opposite helicity sector gives exactly five independent elements (in agreement with the fact that $w$ can appear at most to fourth power to avoid spurious poles in the graviton amplitude),
\begin{align}
(E_3^- \cdot a)^2 (E_4^+ \cdot a)^2&=\frac{(v \cdot \chi)^4}{64 (v \cdot q_{\perp})^4} (w^2-z^2)^2\,,
\nn \\ 
(E_3^- \cdot a) (E_4^+ \cdot a) ((E_3^- \cdot a) (E_4^+ \cdot q_{\perp})+(E_3^- \cdot q_{\perp}) (E_4^+ \cdot a)) a^2 &=- \frac{(v \cdot \chi)^4}{32 (v \cdot q_{\perp})^4} (w^2-z^2) (w -x) z^2\,,
\nn \\ 
(E_3^- \cdot E_4^+) (E_3^- \cdot a) (E_4^+ \cdot a) a^2 &= \frac{(v \cdot \chi)^4}{32 (v \cdot q_{\perp})^4} (w^2-z^2) z^2\,, \nn \\ 
(E_3^- \cdot E_4^+) ((E_3^- \cdot a) (E_4^+ \cdot q_{\perp})+(E_3^- \cdot q_{\perp})) (E_4^+ \cdot a) a^4 &=- \frac{(v \cdot \chi)^4}{16 (v \cdot q_{\perp})^4}  (w -x) z^4\,,
\nn \\ 
(E_3^- \cdot E_4^+)^2 a^4 &= \frac{(v \cdot \chi)^4}{16 (v \cdot q_{\perp})^4} z^4 \,.
\end{align}
Note that the first three vanish for polar scattering, whereas the last two contribute for polar scattering kinematics, $w=\pm z$. This explains our choice in \sec{sect:MatchingR^2} to introduce $3+2$ entire functions encode the complete structure of worldline operators, where only the 3 polar-vanishing ones are relevant for matching the Compton amplitude of \rcite{Cangemi:2023bpe}. 

The analogous classification details holds in the same-helicity sector of spin-2 scattering, there are five independent elementary buliding blocks, matching the allowed powers of $u$. We will not write the formulae, as they are straightforward to obtain from \eqn{eq:3bulidingblocks}.

\bibliographystyle{JHEP}
\bibliography{references}
\end{document}